\documentclass[12pt]{article}

\usepackage{times}
\usepackage{xcolor}
\usepackage{amssymb}
\usepackage{amsmath}
\usepackage{scicite}
\usepackage{graphicx}
\usepackage{subfigure}
\usepackage{threeparttable, longtable, threeparttablex, adjustbox, booktabs, url}

\newenvironment{sciabstract}{%
\begin{quote} \bf}
{\end{quote}}

\title{Frequency Dependent Polarization of Repeating Fast Radio Bursts - Implications for Their Origin}

\author
{Yi Feng,$^{1,2,3}$ Di Li,$^{1,2,12}$$^\ast$ Yuan-Pei Yang,$^{4}$ Yongkun Zhang,$^{1,3} $Weiwei Zhu,$^{1}$ Bing Zhang,$^{5,13}$  \\Wenbin Lu,$^{6}$  Pei Wang,$^{1}$ Shi Dai,$^{7}$ Ryan S. Lynch,$^{8}$ Jumei Yao,$^{9}$ Jinchen Jiang,$^{1,10}$ \\
Jiarui Niu,$^{1,3}$
 Dejiang Zhou,$^{1,3}$ Heng Xu,$^{1,10}$Chenchen Miao,$^{1,3}$  

Chenhui Niu,$^{1}$ \\  Lingqi Meng,$^{1,3}$  Lei Qian,$^{1}$ Chao-Wei Tsai,$^{1}$ 
  Bojun Wang,$^{1,10}$ Mengyao Xue,$^{1}$ \\ 
  Youling Yue,  $^{1}$  Mao Yuan,$^{1,3}$ 
 Songbo Zhang, $^{11}$ Lei Zhang$^{1}$  \\

\normalsize{$^{1}$National Astronomical Observatories, Chinese Academy of Sciences, Beijing 100101, China}\\
\normalsize{$^{2}$Zhejiang Lab, Hangzhou, Zhejiang 311121, China}\\
\normalsize{$^{3}$University of Chinese Academy of Sciences, Beijing 100049, China}\\
\normalsize{$^{4}$South-Western Institute for Astronomy Research, Yunnan University, Kunming, Yunnan, China}\\
\normalsize{$^{5}$ Department of Physics and Astronomy, University of Nevada, Las Vegas, NV 89154, USA}\\
\normalsize{$^{6}$ Department of Astrophysical Sciences, Princeton University, Princeton, NJ 08544, USA}\\
\normalsize{$^{7}$School of Science, Western Sydney University, Penrith NSW 2751, Australia}\\
\normalsize{$^{8}$Green Bank Observatory, Green Bank, WV, 24401, USA}\\
\normalsize{$^{9}$Xinjiang Astronomical Observatory, Chinese Academy of Sciences, Urumqi, Xinjiang 830011, China}\\
\normalsize{$^{10}$School of Physics, Peking University, Beijing 100871}\\
\normalsize{$^{11}$Purple Mountain Observatory, Chinese Academy of Sciences, Nanjing 210023, China}\\
\normalsize{$^{12}$National Astronomical Observatories, Chinese Academy of Sciences-University of KwaZulu-Natal }\\
\normalsize{Computational Astrophysics Centre, University of KwaZulu-Natal, Durban 4000, South Africa}\\
\normalsize{$^{13}$ Nevada Center for Astrophysics, University of Nevada, Las Vegas, NV 89154, USA}\\
\\
\normalsize{$^\ast$ Corresponding author. E-mail:dili@nao.cas.cn} \\

}

\date{}

\begin{document} 


\baselineskip24pt


\maketitle


\begin{sciabstract}
The polarization of fast radio bursts (FRBs), bright astronomical transients,  contains crucial information about their environments.  
 We report polarization measurements of five repeating FRBs, the abundant signals of which enable  wide-band observations with two telescopes. A clear trend of lower polarization at lower frequencies was found, which can be well characterized by a single parameter rotation-measure-scatter ($\sigma_{\mathrm{RM}}$) and modeled by multi-path scatter. 
 Sources with higher $\sigma_{\mathrm{RM}}$ have higher RM magnitude and scattering timescales. The two sources with the most substantial $\sigma_{\mathrm{RM}}$, FRB~20121102A and FRB~20190520B, are associated with a compact persistent radio source. 
 These properties indicate a complex environment near the repeating FRBs, such as a supernova remnant or a pulsar wind nebula, consistent with their arising from young  populations.

\end{sciabstract}



Fast radio bursts (FRBs) are the brightest millisecond-duration-astronomical transients in radio bands\cite{2007Sci...318..777L}. Some FRBs are known to repeat, such as FRB~20121102A \cite{2016Natur.531..202S}, but it remains unclear whether repeating fast radio bursts (hereafter repeaters) are ubiquitous or peculiar sources. Whether all FRBs repeat on some time scale, or if repeaters form a separate population from single-burst sources, has implications for the energy source \cite{2020Natur.587...45Z}, host environments \cite{2017ApJ...834L...8M}, luminosity function \cite{2018MNRAS.481.2320L}, apparent and intrinsic event rates \cite{2019A&A...632A.125G}, and predicted survey yields \cite{2018MNRAS.474.1900M}. 

Information about the FRB hosts and environments could be obtained from their polarization. Faraday rotation is a phenomenon where the polarization position angle of linearly polarized radiation propagating through
a magneto-ionic medium is rotated as a function of frequency, quantified by the Rotation Measure (RM), which carries information about the local environment and the intervening medium. The substantial and time-varying RM of FRB~20121102A has been interpreted as due to a dynamic magneto-ionic environment\cite{121102rm}, such as an expanding supernova remnant or a pulsar wind nebula\cite{2018ApJ...861..150P,2018ApJ...868L...4M}. The polarization position angle and fraction of linear/circular polarization could constrain the emission mechanisms\cite{2019MNRAS.483..359L,dai20}. For example, a constant polarization position angle across a burst is consistent with either a far-away model invoking a relativistic shock or emission from the outer magnetosphere of a neutron star \cite{2020Natur.587...45Z}. Alternatively, the varying polarization angle observed in the repeater FRB~20180301A has been interpreted as originating within the magnetosphere of a magnetar (a highly magnetized neutron star) \cite{luo2020}.

We analyze the polarization properties of 21 FRB sources, 9 of which have been confirmed to repeat. We observed five of them, FRB~20121102A, FRB~20190520B, FRB~20190303A,  FRB~20190417A, and FRB~20201124A using the Five-hundred-meter Aperture Spherical radio Telescope (FAST) and the Robert C. Byrd Green Bank Telescope (GBT). The other 16 sources are taken from the literature (listed in Table S1). Our goal is to examine their RM, linear polarization fraction and its dependence on frequency.

The repeater FRB~20121102A has been precisely localized to a dwarf galaxy\cite{2016Natur.531..202S,2017Natur.541...58C}. It has a bimodal burst energy distribution with 1652 independent bursts detected in 59.5 hours spanning 62 days \cite{li21}. Using the same dataset, we searched RM in each of the 1652 bursts and detected no linear polarization setting a 6\%  upper limit on the degree of linear polarization  at 1.0-1.5\,GHz\cite{supmm}. For comparison, previous observations showed almost 100\% linear polarization at 3-8\,GHz\cite{121102rm}. 

\begin{figure}[!htp]
\centering
\subfigure[FRB~20190303A]{
\includegraphics[width=0.45\textwidth]{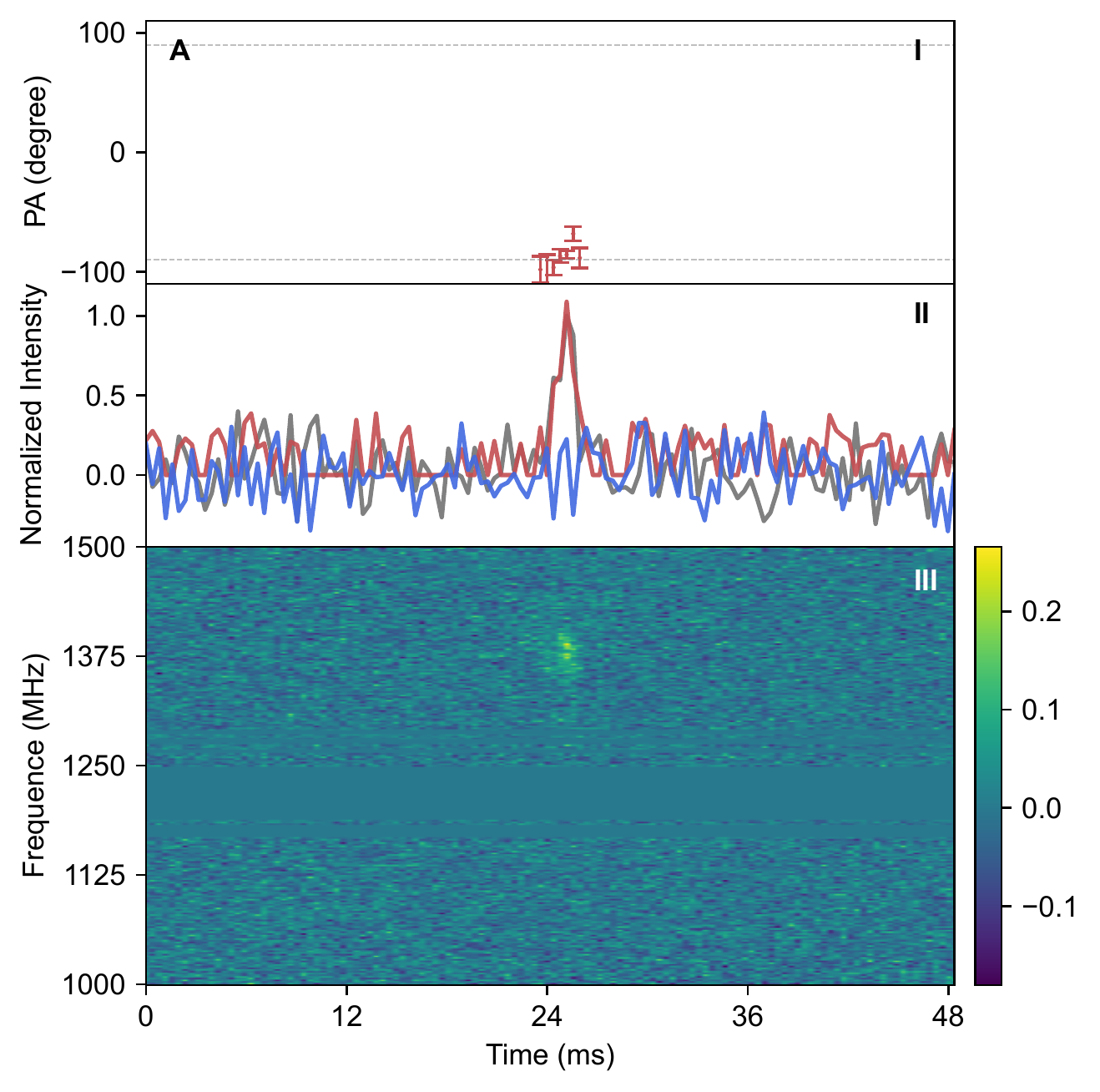}}
\subfigure[FRB~20190417A]{
\includegraphics[width=0.45\textwidth]{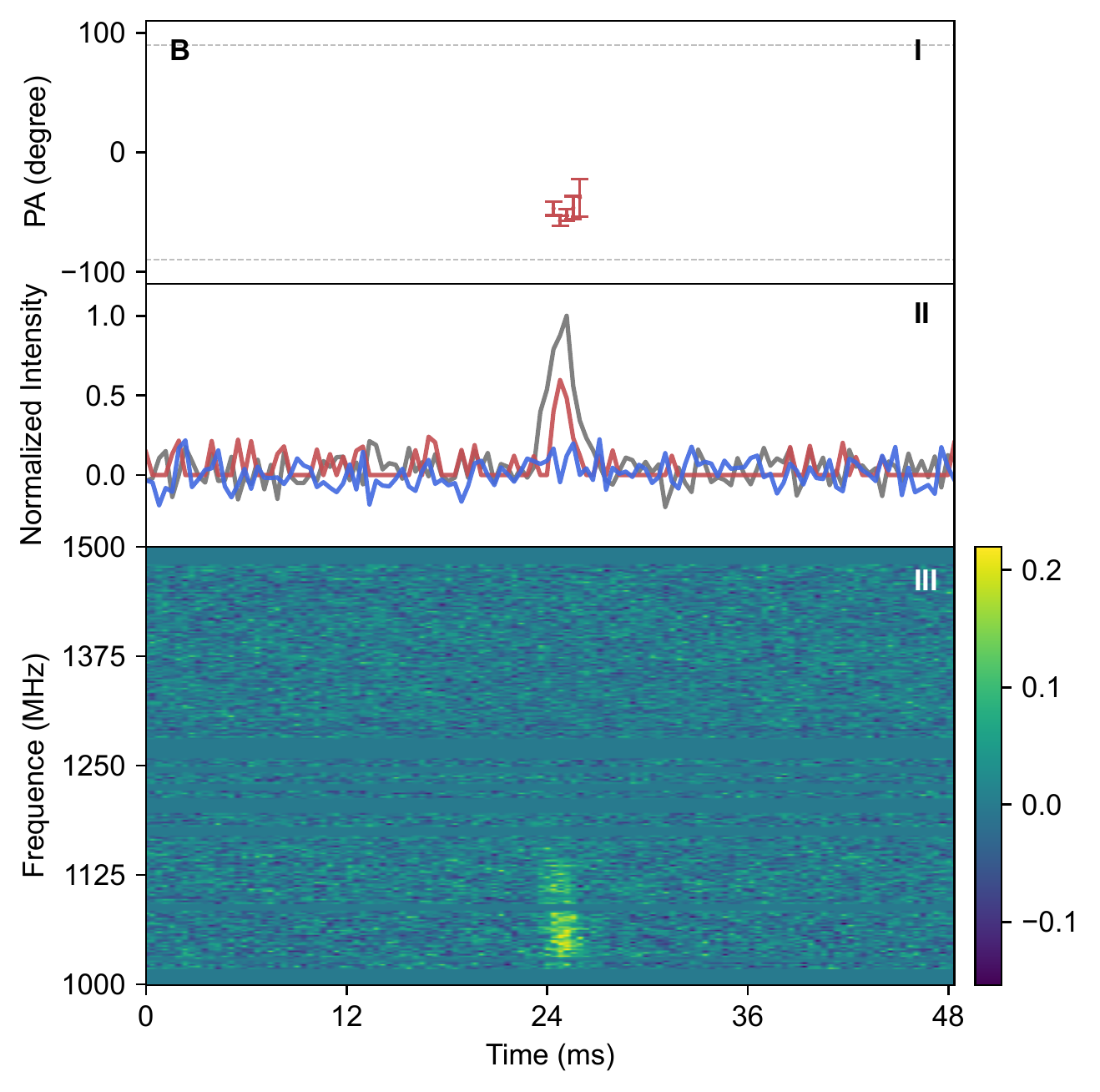}}\\
\subfigure[FRB~20190520B]{
\includegraphics[width=0.45\textwidth]{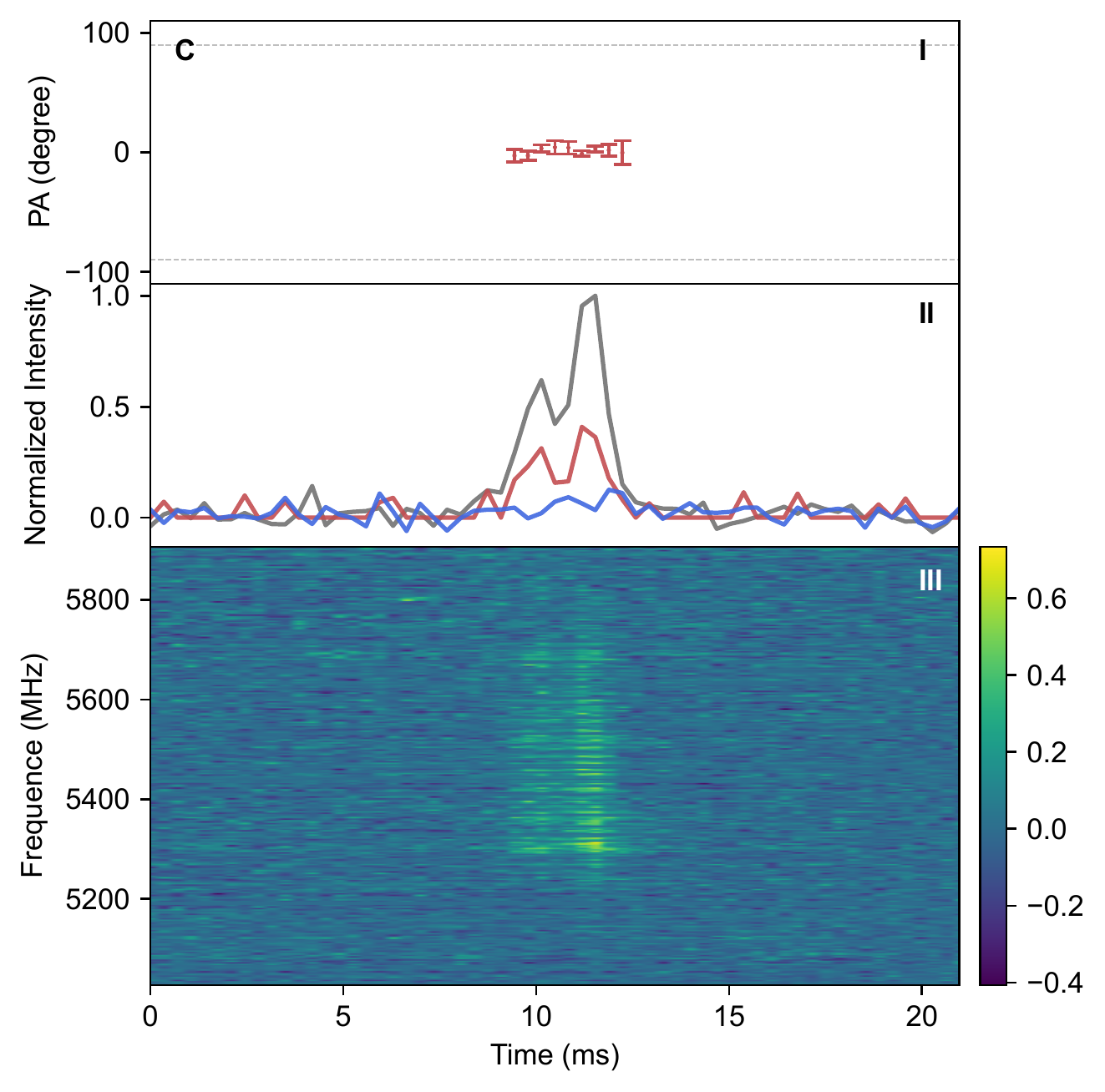}}
\subfigure[FRB~20201124A]{
\includegraphics[width=0.45\textwidth]{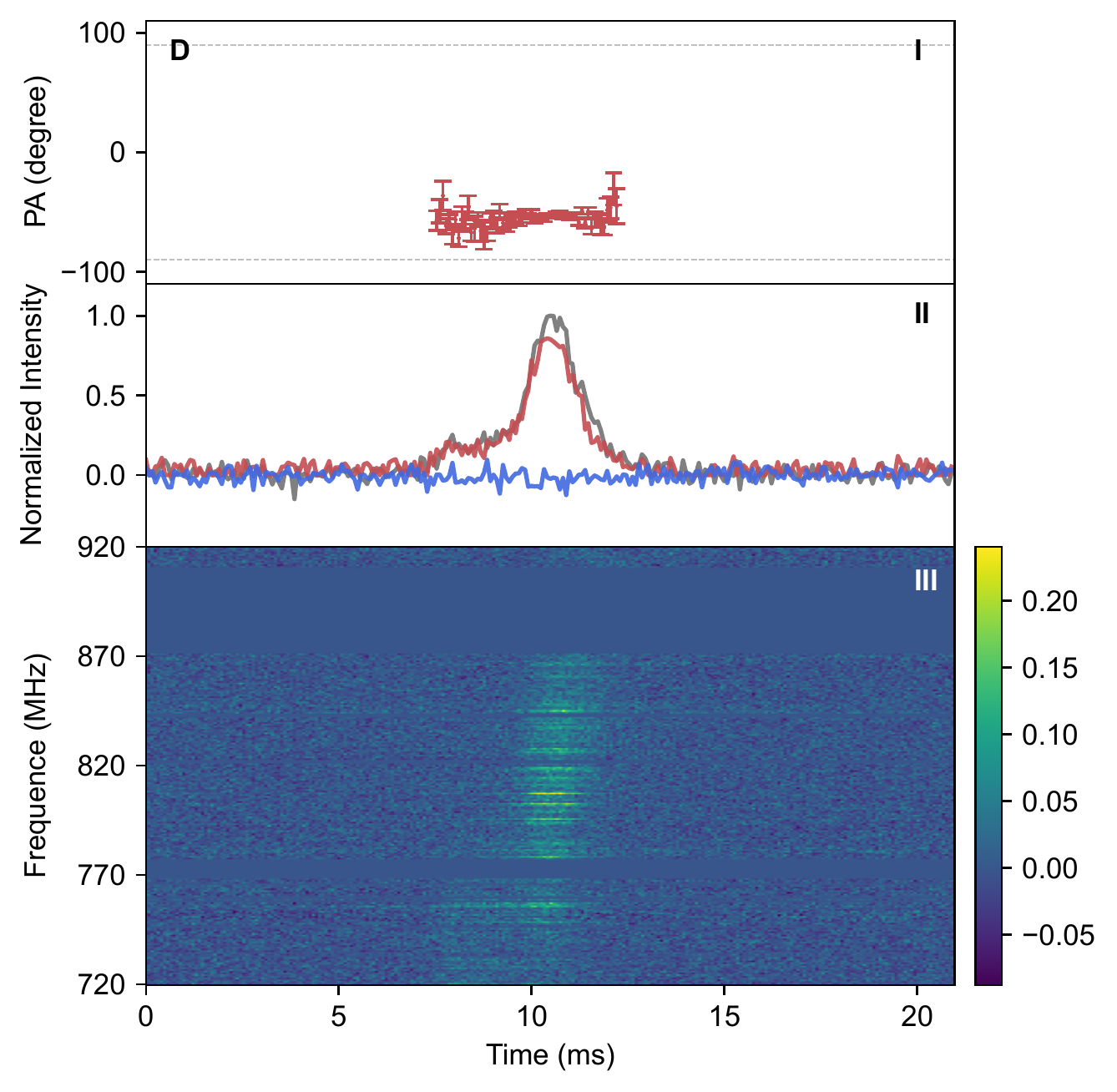}}

\caption{\textbf{Polarization position angle and intensity profiles with dynamic spectra for four repeating FRBs.}
In each panel, sub-panel I shows the polarization position angles with gray dashed lines indicating $\pm90^{\circ}$; sub-panel II shows the polarization pulse profiles with lines indicating total intensity (black, normalized to a peak value of 1.0), linear polarization (red) and circular polarization (blue); sub-panel III shows the dynamic spectra.}
\label{fig:pulse}
\end{figure}

FRB~20190520B is another repeater localized to a persistent radio source (PRS) in another dwarf galaxy \cite{li18,niu21}. An RM search for the 75 bursts presented in Ref.~18 resulted in no detection, with an upper limit of 20\% on the degree of linear polarization at 1.0-1.5\,GHz \cite{supmm}. Our follow-up observations with the GBT detected three bursts from 4-8\,GHz band with an average RM of 2759\,rad m$^{-2}$ \cite{supmm}. Representative polarization pulse profiles and dynamic spectra of this source are shown in Figure~\ref{fig:pulse}C.

FRBs~20190303A, 20190417A, and 20201124A were discovered at 400–800\,MHz  \cite{chime9,2021ATel14497....1C} . We followed up these sources with the FAST 19-beam system at 1.0-1.5\,GHz. From FRB~20190303A, 3 bursts were detected with an average RM of -411\,rad m$^{-2}$, which help localize the source to 
(Right Ascension--RA, Declination--Dec) = (13$^h$51$^m$58$^s$,  +48$^\circ$07$'$20$''$) (J2000) with a 2.6$'$ error circle \cite{supmm}. Compared with the earlier lower-frequency measurements\cite{chime9}, the RM has changed about 100\,$\mathrm{rad/m}^2$ over 1.5-year timespan. In contrast to the earlier low linear polarization\cite{chime9}, the bursts detected with FAST are nearly 100\% linearly polarized. From FRB~20190417A, we detected 23 bursts, 5 of which have measurable polarization  with an average RM of 4681\,rad m$^{-2}$. The estimated position of FRB~20190417A is 
(RA, DEC) = (19$^h$39$^m$22$^s$, +59$^\circ$18$'$58$''$) (J2000) with an error circle of 2.6$'$ \cite{supmm}. FAST follow-up observations of FRB~20201124A resulted in a large sample bursts, of which 11 were bright enough to be used in our analysis \cite{supmm}. We also detected 9 polarized bursts with the GBT at 720–920\,MHz, with an average RM of -684\,rad m$^{-2}$. Both FRB~20190417A and FRB~20201124A have higher linear polarization at higher frequencies. The average RMs of each FRB are listed in Table~S2. 
The time of arrival (ToA), the central frequency of the burst emission (weighted by the burst signal-to-noise ratio as a function of frequency), the intra-channel depolarization $f_{\mathrm{depol}}$ (see below and Eq.~\ref{eq:depol}), RM, and degree of de-biased \cite{supmm} linear and circular polarization of each pulse are listed in Table~S3. Representative polarization pulse profiles and their dynamic spectra are shown in Figure~\ref{fig:pulse} \cite{supmm}.

The observations of these four sources all indicate decreasing linear polarization with decreasing frequencies. To explain this behaviour, we consider three  effects: i) intrinsic frequency evolution of the linear polarization, ii) intra-channel depolarization, and iii) RM scatter. 

Intrinsic frequency evolution of linear polarization is already known for pulsars \cite{2018MNRAS.474.4629J}. The polarization tends to decrease from lower to higher frequencies, which has been attributed to emission from different heights in the pulsar magnetospheres \cite{2001A&A...378..883P}. This is the opposite trend to the one we find for repeating FRBs, so a direct analogy is not supported. Given the lack of understanding of FRB origin(s), our results cannot rule out other scenarios involving pulsar-magnetosphere-like environments. 

The alternative intra-channel depolarization $f_{\rm depol}$, the fractional reduction in the linear polarization amplitude, is defined as: 
\begin{equation}
\label{eq:depol}
\begin{split}
f_{\rm depol} \equiv 1-\frac{{\rm sin}(\Delta{\theta})}{\Delta{\theta}},\\
\end{split}
\end{equation}
where $\Delta{\theta}$ is the intra-channel polarization position angle rotation. $\Delta{\theta}$ is defined as $\Delta\theta \equiv \frac{{2\rm RM}_{\rm obs}c^2\Delta\nu}{\nu_c^{3}}$, where $\rm RM_{\rm obs}$ is the observed rotation measure, $c$ is the speed of light, $\Delta{\nu}$ is the channel frequency width, and $\nu_c$ is the central channel observing frequency. 
For repeaters, the measured RMs are too small for explaining the depolarization through this effect ($f_{\rm depol}$ listed in Table~S1 and Table~S3). We infer that intra-channel depolarization is unlikely to be a major cause of depolarization for repeating FRBs. 
\begin{figure}[!htp]
\centering
\includegraphics[width=0.9\textwidth]{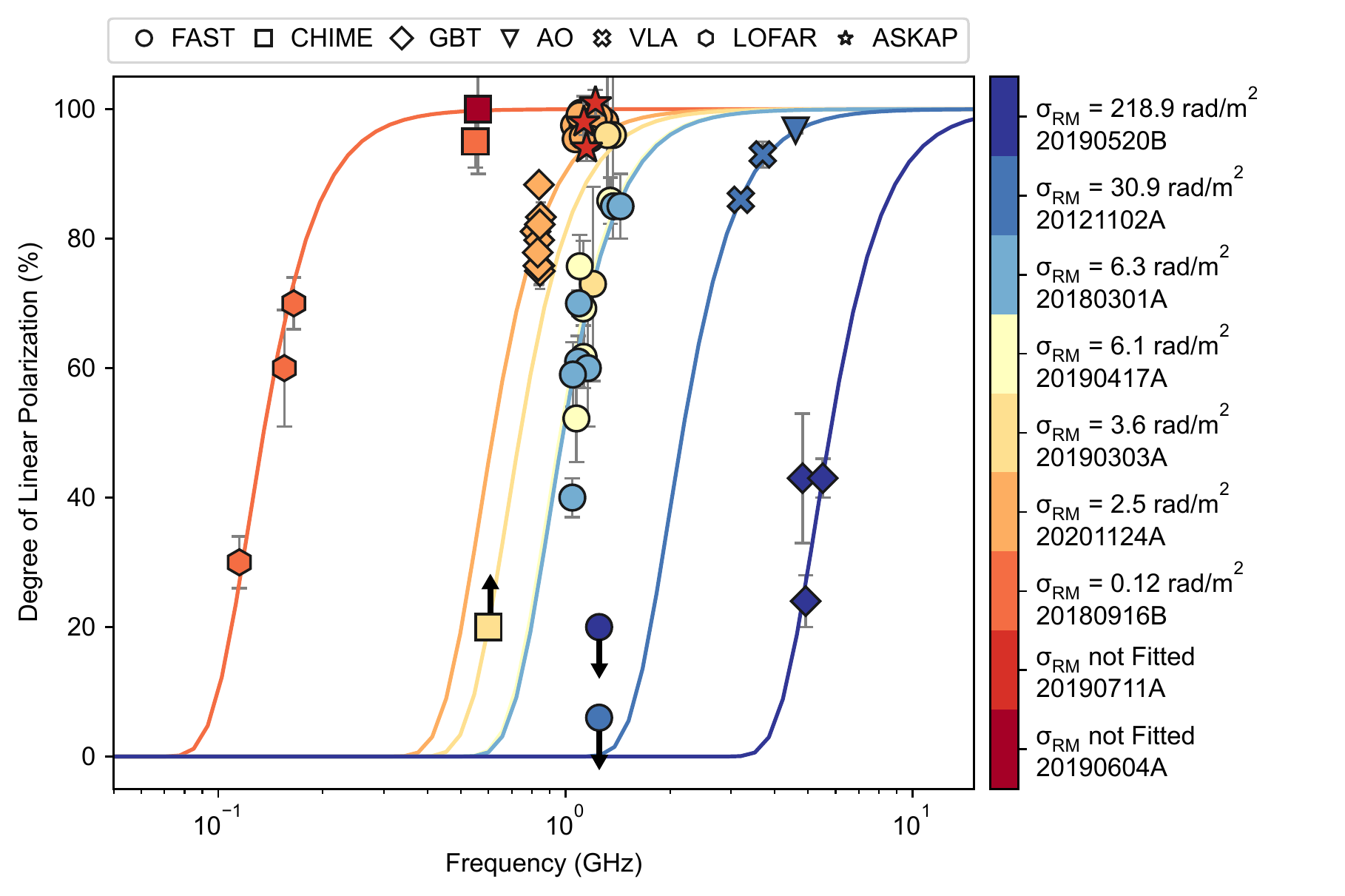}
\caption{\textbf{The degree of linear polarization for FRB sources is consistent with RM scattering.} Data points with $1\sigma$ error bars indicate the degree of linear polarization as a function of frequency for each FRB. The colored lines are models of emission that is intrinsically 100\% linearly polarized then depolarized by various $\sigma_{\mathrm{RM}}$ levels (Equation 2), fitted to each FRB separately. Arrows indicate $95\%$ upper and lower limits. All the bursts in the sample are consistent with the RM scattering models. Data values and sources are listed in Tables S1 and S3, and the model $\sigma_{\rm RM}$ values in Table S2. CHIME, AO, VLA, LOFAR and ASKAP at the top represents Canadian Hydrogen Intensity Mapping Experiment, Arecibo Observatory, Karl G. Jansky Very Large Array, Low-Frequency Array and Australian Square Kilometre Array
Pathfinder, respectively.} 
\label{fig:rmscatter}
\end{figure}

\begin{figure}[!htp]
\centering
\includegraphics[width=0.9\textwidth]{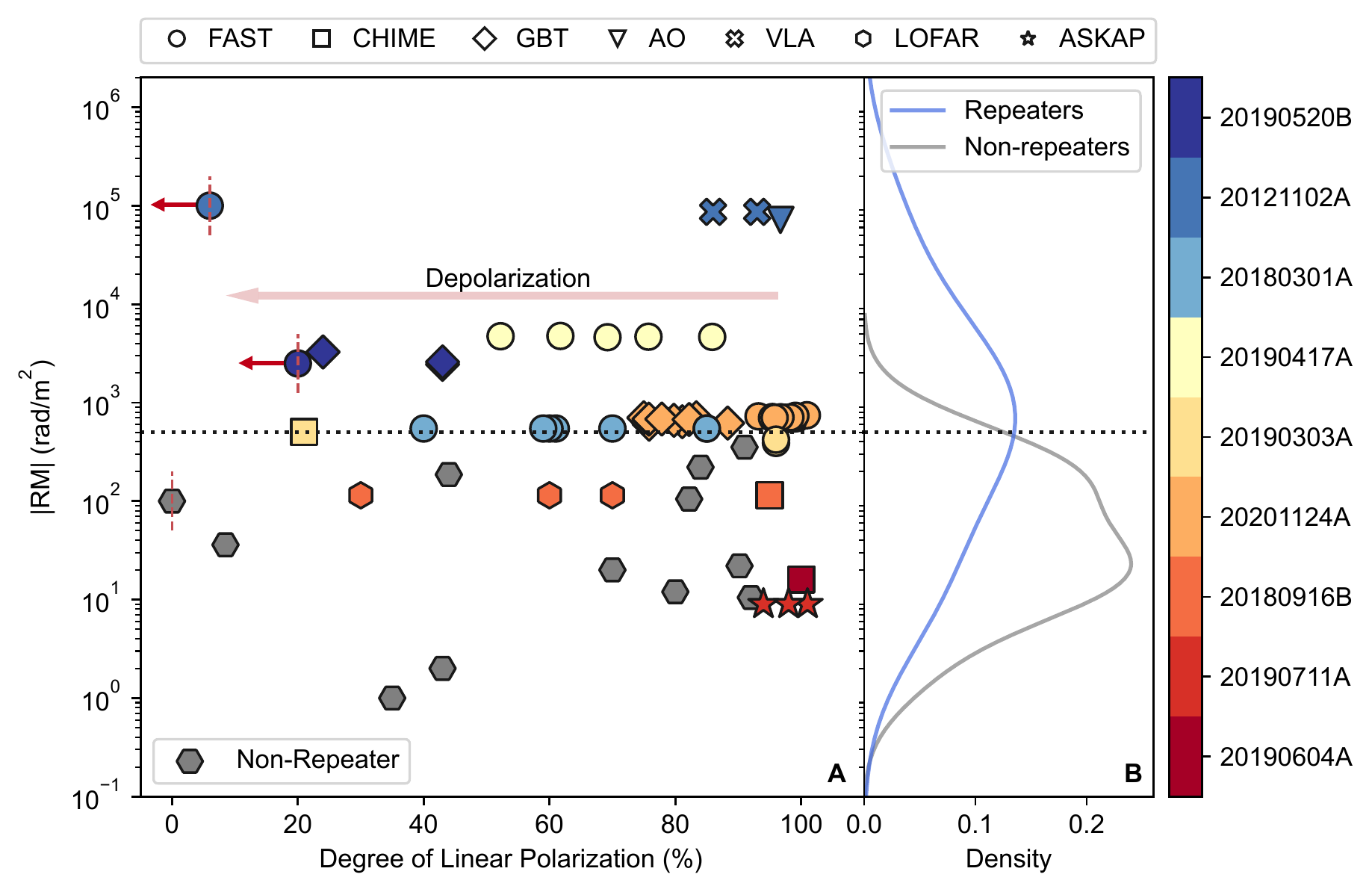}
\caption{\textbf{The relation between RM and degree of linear polarization for repeating and non-repeating FRBs.} The vertical dashed red lines indicate that the respective RM was not detected in L-band. We thus choose nominal values that are either representative or measured at higher frequencies, i.e.\ 100\,rad m$^{-2}$ for FRB~20140514A, $3000$\,rad m$^{-2}$ for FRB~20190520B and $10^5$\,rad m$^{-2}$ for FRB~20121102A.  The red arrows are $95\%$ upper limits on the linear polarization. The horizontal dotted black line indicates $|\mathrm{RM}|=500\, \mathrm{rad/m}^2$, which lies above all the non-repeaters. The blue and gray profiles in the panel B are the Kernel Density Estimation (KDE) of the repeaters' RMs and non-repeaters' RMs. The p-value from Kolmogorov-Smirnov test of the repeaters' RM and non-repeaters' RM is 0.02. The acronyms of telescopes at the top are the same as those in Figure \ref{fig:rmscatter}.}
\label{fig:L-RM}
\end{figure}
We next examine RM scattering, the dispersion of RM about the apparent mean for each source \cite{2012MNRAS.421.3300O}. RM scattering can be caused by multi-path transmission of signals in an inhomogeneous magneto-ionic environment. If the scattering is large enough, it can become substantial enough to depolarize the pulses, analogous to pulsar pulses passing through a stellar wind\cite{2018ApJ...867...22Y,2019MNRAS.490..889P}. We parametrize the depolarization due to RM scattering as \cite{2012MNRAS.421.3300O}:
\begin{equation}
\label{eq:rmscatter}
f_{\rm{RM~scattering}} \equiv 1 - \exp{(-2\lambda^4\sigma^2_{\mathrm{RM}})} \>\>,
\end{equation}
where $f_{\rm{RM~scattering}}$ is the fractional reduction in the linear polarization amplitude, $\sigma_{\mathrm{RM}}$ is the standard deviation of the RM and $\lambda$ is the wavelength.

Depolarization at lower frequencies, consistent with irregular RM variations, has been seen in a few pulsars with scatter broadening. For example, the variable degree of linear polarization observed in the pulsar PSR~J0742$-$2822 between 200\,MHz and 1\,GHz can be described by Eq.~\ref{eq:rmscatter} with $\sigma_{\mathrm{RM}} = 0.13~ \mathrm{rad/m}^2$ \cite{Xue2019}.

Complex magneto-ionic environments have previously been inferred from observations for some repeaters\cite{121102rm,luo2020}. We also expect RM scatter for FRBs in such environments. In Figure \ref{fig:rmscatter}, we show the degree of linear polarization versus frequency for each source individually. Frequency evolution can be seen for all sources, but the depolarization occurs at different bands. 
FRB~20180916B is depolarized below 200 MHz\cite{lofar21}, while FRB~20121102A\cite{2021ApJ...908L..10H}  is depolarized at frequencies lower than 3.5\,GHz. 
We fitted the data for each repeater using the model in Equation 2, with a different $\sigma_{\mathrm{RM}}$ for each source \cite{supmm}. For the sources depolarized at lower frequencies ($<$200 MHz), such as FRB~20180916B, we derive $\sigma_{\mathrm{RM}} \sim0.1~\mathrm{rad/m}^2$ (Table S2).  
Such a small scatter is consistent with the environment being less turbulent, dense, and/or magnetized than that of other FRBs, as might be expected for an old stellar population \cite{lofar21}. We find $\sigma_{\mathrm{RM}} \sim2.5\, \mathrm{rad/m}^2$ for FRB~20201124A, $\sigma_{\mathrm{RM}} \sim3.6\, \mathrm{rad/m}^2$ for FRB~20190303A, $\sigma_{\mathrm{RM}} \sim6.1\,\mathrm{rad/m}^2$ for FRB~20190417A, $\sigma_{\mathrm{RM}} \sim6.3\, \mathrm{rad/m}^2$ for FRB~20180301A and $\sigma_{\mathrm{RM}} \sim30.9\,\mathrm{rad/m}^2$ for FRB~20121102A (Table S2). The RM scatter of 30.9\,$\mathrm{rad/m}^2$ is consistent with FRB~20121102A originating in a younger stellar population \cite{2021ApJ...908L..10H} that produced an inhomogeneous magneto-ionic environment.
For the repeater FRB~20190520B, which depolarizes at frequencies higher than 1\,GHz, we derive $\sigma_{\mathrm{RM}} \sim218.9\,\mathrm{rad/m}^2$. The observations are consistent with repeating bursts being intrinsically nearly 100\%  linearly polarized, then being depolarized through transmission processes, which can be  characterized by the RM scatter parameter $\sigma_{\mathrm{RM}}$. The same environment that gives rise to large $\sigma_{\mathrm{RM}}$ could also cause Faraday conversion wherein linearly polarised light is converted to circularly polarised light, potentially explaining the circular polarization observed in some FRBs \cite{2021MNRAS.508.5354H}.
We interpret the $\sigma_{\mathrm{RM}}$ values derived from our analysis as a measure or the complexity level of the magneto-ionic environments hosting active repeaters, with larger $\sigma_{\mathrm{RM}}$ values being possibly associated with younger populations.

Figure~\ref{fig:L-RM} shows the relation between RM and linear polarization for repeating and non-repeating FRBs in our sample with polarization measurements. 
A Kolmogorov-Smirnov test between the repeaters and non-repeaters finds the RM distributions differ, indicating they may reside in different environments. We caution that the intra-channel depolarization may introduce a selection bias if the non-repeaters have systematically larger RMs, because discovery searches tend to use coarser filter banks than follow-up observations.

\begin{figure}[!htp]
\centering
\includegraphics[width=0.75\textwidth]{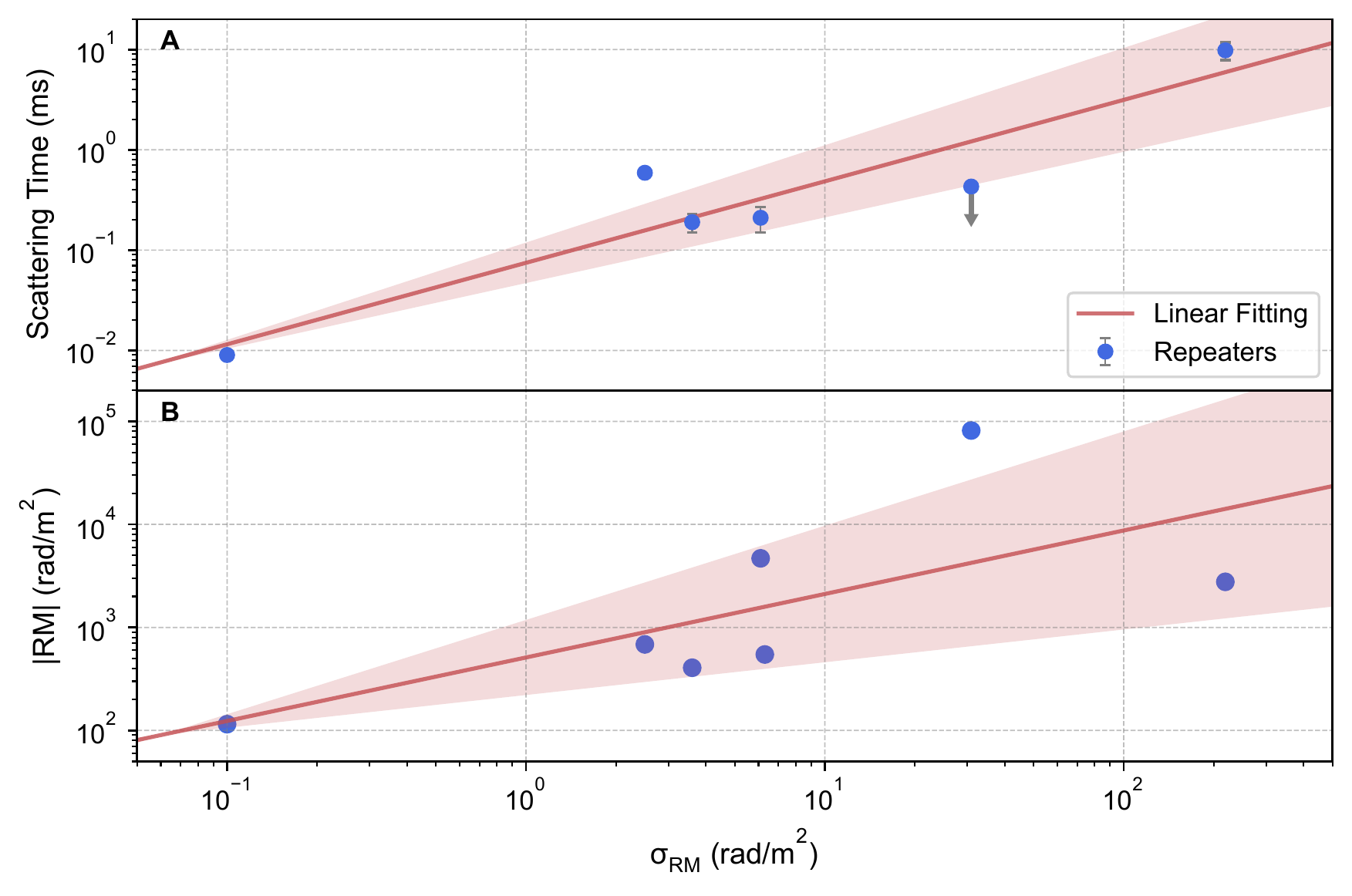}
\caption{ \textbf{Correlations between $\sigma_{\mathrm{RM}}$, scattering time $\tau_{\mathrm{ sca}}$ and rotation measure magnitude $|$RM$|$ for repeating FRBs.} Panel A: The relation between $\sigma_{\mathrm{RM}}$ and the scattering timescale scaled to 1300\,MHz\cite{supmm} for repeaters with $\sigma_{\mathrm{RM}}$ measurements. The data used are listed in Table~S2. FRB~20121102A is an upper limit (gray arrow). The Pearson product-moment correlation coefficient of $\log\tau_{\mathrm{ sca}}$ and $\log\sigma_{\mathrm{RM}}$ is 0.47. Fitting the data with a linear model (red line) gives a slope of $0.81 \pm 0.16$. Panel B: The relation between $\sigma_{\mathrm{RM}}$ and $|$RM$|$ for the same sample of repeaters. The Pearson product-moment correlation coefficient of $\log \mathrm{|RM|}$ and $\log\sigma_{\mathrm{RM}}$ is 0.68. Fitting the data with a linear model (red line) gives a slope of $0.62\pm0.30$. All quantities are measured in the observer's frame. The $1\sigma$ uncertainties on $\sigma_{\mathrm{RM}}$ and $|$RM$|$ are smaller than the symbol size.}
\label{fig:rm_rms}
\end{figure}

If $\sigma_{\mathrm{RM}}$ is due to multi-path propagation effect in individual FRB bursts, we would expect the same effect to produce a temporal scattering in the pulse profile. 
In Figure~\ref{fig:rm_rms}, we show RM scatter and scattering timescales, which are listed in Table~S2, for our repeater sample. The two are positively correlated. 
This is consistent with the hypothesis that RM scatter and pulse scattering originate from a single plasma screen, such as a supernova remnant or pulsar wind nebula \cite{supst}. In Figure~\ref{fig:rm_rms},  $\sigma_{\mathrm{RM}}$ and $\rm |RM|$ are shown to be correlated positively, which, in the context of our multi-path-scattering model, indicate an environment, where a stronger magnetic field strength $B$ tend to have a larger fluctuation in $B$. 

Only two FRBs FRB~20190520B and FRB~20121102A, are known to have associated compact PRSs \cite{niu21,2017ApJ...834L...8M}. We find that these two repeaters have the largest $\sigma_{\mathrm{RM}}$. The combination of large RM (dense) and strong $B$ field (magnetized) tend to produce large $\sigma_{\mathrm{RM}}$.  A denser and more magnetized environment likely also produces stronger synchrotron radiation from the nebula, resulting in a PRS \cite{2020ApJ...895....7Y}, consistent with the multi-path-scattering picture. Repeaters with  large observed $\sigma_{\mathrm{RM}}$ could be more affected by turbulence, resulting in large fluctuations of electron density and magnetic field, which may explain the diversity among repeaters \cite{supst}. 

In summary, $\sigma_{\mathrm{RM}}$ can be used to quantify the complexity of the magnetized environments, which seem to be commonly associated with repeating FRBs. The more substantial the value of  $\sigma_{\mathrm{RM}}$ 
 possibly indicates the source being younger. 

\bibliographystyle{Science}

\begin{thebibliography}{10}
\bibitem{supmm} The observations and data analysis are described in Materials and Methods.
\end{thebibliography}

\section*{Acknowledgments}
We thank Brenne Gregory for help with the observations and Emmanuel Fonseca, G. H. Hilmarsson, Daniele Michilli, Jason Hessels, Adam Deller, Cherie Day, Steve White and Bob Simon for valuable discussions. This work made use of data from FAST, a Chinese national mega-science facility, operated by National Astronomical Observatories, Chinese Academy of Sciences. The Green Bank Observatory is a facility of the National Science Foundation operated under cooperative agreement by Associated Universities, Inc.  

\section*{Funding}
D.L. and Y.F. are supported by NSFC grant No. 11988101, 11725313, 11690024, by the National Key R$\&$D Program of China No. 2017YFA0402600, and by Key Research Project of Zhejiang Lab No. 2021PE0AC03. Y.P.Y is supported by NSFC grant No. 12003028. W.W.Z. is supported by National SKA Program of China No.2020SKA0120200 and the NSFC 12041303, 11873067. W.B.L. is supported by Lyman Spitzer, Jr. Fellowship at Princeton University. P.W. is supported by NSFC grant No. U2031117, the Youth Innovation Promotion Association CAS (id.~2021055), CAS Project for Young Scientists in Basic Reasearch (grant YSBR-006) and the Cultivation Project for FAST Scientific Payoff and Research Achievement of CAMS-CAS. S. D. is the recipient of an Australian Research Council Discovery Early Career Award (DE210101738) funded by the Australian Government. J.M.Y. is supported by NSFC grant No. 11903049, and the Cultivation Project for FAST Scientific Payoff and Research Achievement of CAMS-CAS. L.Q. is supported by National SKA Program of China No. 2020SKA0120100 and the Youth Innovation Promotion Association of CAS (id.~2018075). S.B.Z. is supported by NSFC grant No. 12041306, and by China Postdoctoral Science Foundation (Grant No. 2020M681758). L.Z. is supported by NSFC grant No. 12103069. 

\section*{Author Contributions}
Y.F. and D.L. developed the concept of the manuscript. Y.F. and D.L. led the discussion on the interpretation of the results and writing of the manuscript. Y.P.Y., B.Z. and W.B.L. derived the theoretical formula. Y.F., D.L., W.W.Z., B.Z., W.B.L., R.S.L. and C.H.N. arranged the observations. Y.F. and Y.K.Z. conducted the main data analysis and visualization. All authors contributed to the analysis or interpretation of the data and to the final version of the manuscript.     
\section*{Competing interests}
The authors declare no competing interests.

\section*{Data and materials availability}
Observational data of FAST (Project ID: ZD2020\_5) are available from the FAST archive (\url{http://fast.bao.ac.cn}) 1 year after data collection, following FAST data policy. Observational data of GBT (Project ID: 21A-404 and 21A-416) are available from the GBT archive (\url{https://greenbankobservatory.org}) 1 year after data collection, following GBT data policy. 
The archival data sources are listed in Table S1. The numerical results of the analysis are listed in Tables S2 and S3. The bursts data for Tables S2 and S3 are openly available in Science Data Bank at \url{https://doi.org/10.11922/sciencedb.o00069.00006}. Computational programs for the polarization analysis reported here are available at \url{https://github.com/SukiYume/RMS}. Other standard data reduction packages are available at their respective websites: \textsc{presto} (\url{https://github.com/scottransom/presto}), \textsc{heimdall} (\url{http://sourceforge.net/projects/heimdall-astro/}), \textsc{psrchive} (\url{http://psrchive.sourceforge.net}).

\section*{Supplementary materials}
Materials and Methods\\
Supplementary Text\\
Figs.\ S1 to S5\\
Tables.\ S1 to S3\\
References \textit{(32-61)}


\clearpage

\end{document}


\maketitle 


\baselineskip24pt



\setcounter{section}{0}
\renewcommand{\thesection}{S\arabic{section}}
\setcounter{equation}{0}
\renewcommand{\theequation}{S\arabic{equation}}
\setcounter{figure}{0}
\renewcommand{\thefigure}{S\arabic{figure}}
\setcounter{table}{0}
\renewcommand{\thetable}{S\arabic{table}}
\setcounter{subsection}{1}

\section*{This PDF file includes:}
Materials and Methods\\
Supplementary Text\\
Figs.\ S1 to S5\\
Tables\ S1 to S3\\

\clearpage

\paragraph{\bf\LARGE Materials and Methods}
\section{Observations and burst detection}
\subsection{FRB~20190520B}

FRB~20190520B was observed from 3.95--7.8 GHz with the GBT's C-Band receiver and the Versatile Green Bank Astronomical Spectrometer (VEGAS) digital backend\cite{2015ursi.confE...4P}.  VEGAS consists of eight spectrometer banks, each of which can sample 1.5 GHz of bandwidth, though filters reduce the usable bandwidth to 1.25 GHz per bank.  In the case of FRB~20190520B, we used four VEGAS banks centered on 4312.5, 5437.5, 6562.5, and 7687.5 GHz.  Each of these sub-bands overlapped the next by 187.5 MHz, ensuring complete sampling over the frequencies of interest.  The data were then combined in post-processing to cover the full available receiver band.

Data were recorded in the \textsc{psrfits} standard format \cite{2004PASA...21..302H}. Full-Stokes spectra were recorded every 43.907 $\upmu$s with 0.366 MHz-wide channels (i.e., 4096 channels per spectrometer bank).

We searched for bursts with widths up to 10 ms in dedispersed time series using a matched filtering algorithm as implemented by the \textsc{presto} program \cite{2001PhDT.......123R} \textsc{single\_pulse\_search.py}.  We retained the native sampling rate of the full-resolution data but reduced the frequency resolution by a factor of four to increase computational efficiency.  Using wider channel bandwidths leads to slightly more intra-channel dispersive smearing, but the effect is not large.  For example, at the lowest frequency in our observing band (3.95 GHz) the dispersive smearing at our native frequency resolution is 0.06 ms, and in the reduced-resolution data it is 0.2 ms, which is still much less than the typical FRB pulse width. The data were then dedispersed using the \textsc{presto} program \textsc{prepsubband} at dispersion measures (DMs) ranging from 1110--1309 pc~cm$^{-3}$ with a step size of 1 pc~cm$^{-3}$.  Searching for bursts over a range of DMs allowed us to differentiate between astrophysical sources and radio frequency interference by looking for a characteristic peak in signal to noise ratio (S/N) near the true DM of the FRB and a monotonically decreasing S/N as the error in DM increases.  We examined all candidate bursts with S/N$>$6 to confirm or reject their astrophysical nature.

Data were calibrated using the \textsc{psrchive} package \cite{2004PASA...21..302H}.  On and off-source observations of the standard calibrator 3C 394 were used to measure the intensity of the C-Band receiver's built-in noise diode.  The noise diode was observed again at the position of FRB~20190520B prior to the main observations, and these data were used to calibrate each burst from the FRB.

\subsection{FRB~20190303A}
Observations of FRB~20190303A were carried out using the FAST telescope mounted with the 19-beam receiver\cite{2019SCPMA..6259502J}, which operates with a frequency range from 1050-1450\,MHz and provides two data streams (one for each linear polarization). The data streams are processed with the Reconfigurable Open Architecture Computing Hardware–version 2 (ROACH2) signal processor\cite{2019SCPMA..6259502J}. The output data files are recorded as 8 bit-sampled search mode \textsc{psrfits} files with 4096 frequency channels.

\begin{figure}[!htp]
\centering
\includegraphics[width=0.9\textwidth]{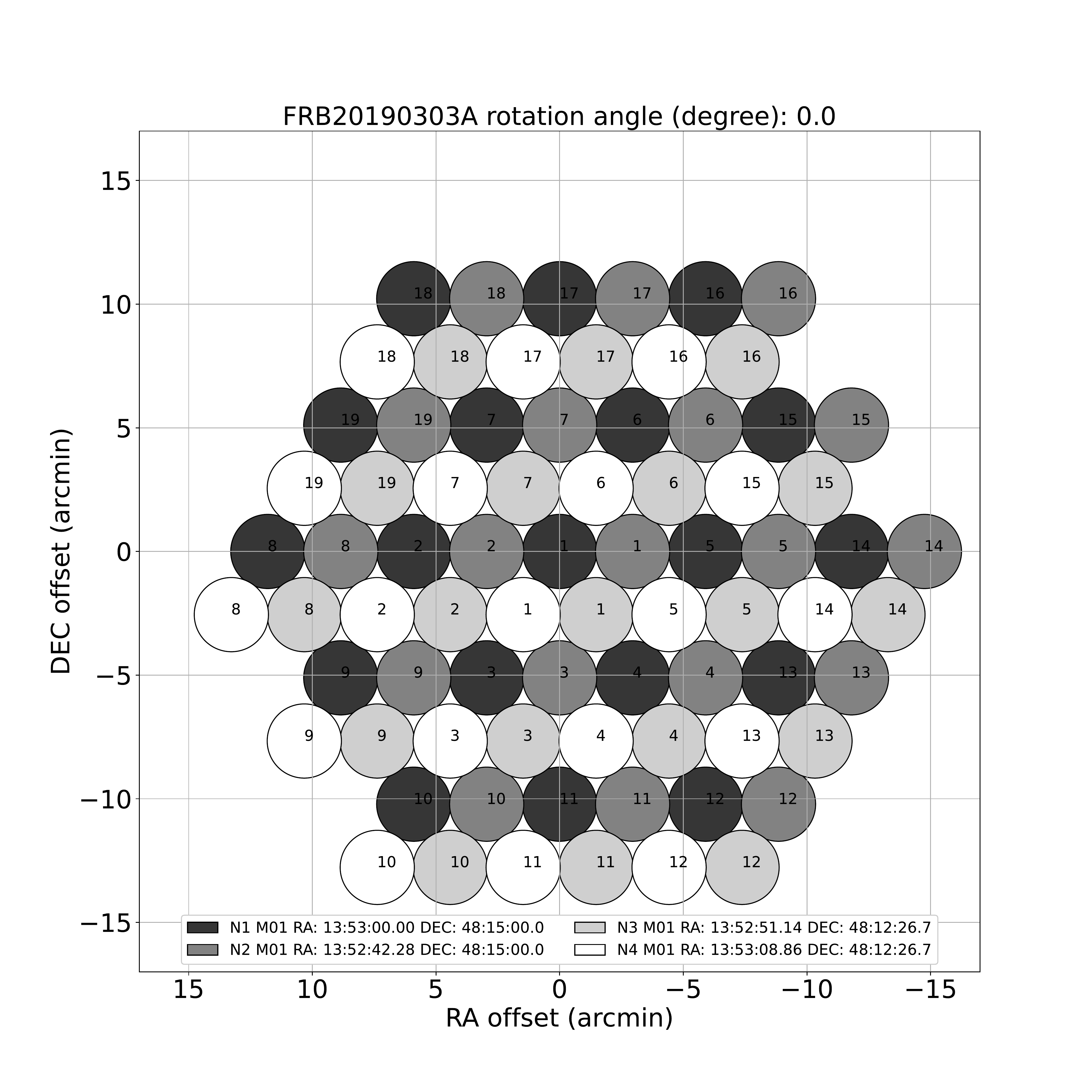}
\caption{\textbf{Gridding observations.} Gridding observations of FRB~20190303A on 4 Jan 2021 (N1, N2) in legend and 5 Jan 2021 (N3, N4). Each pointing of N1, N2, N3 and N4 lasts for 30 minutes. }
\label{fig:grid}
\end{figure}

Observations were carried out in six sessions. We carried out the first and second session on 4 Jan 2021 and 5 Jan 2021 to do a 2-hour gridding observation using all beams of the 19-beam receiver. 196.608\,$\upmu$s time resolution was used. The central beam of the receiver was initially placed on the previously reported position (RA = 13h53min, Dec = +48$^\circ$15$'$) (J2000)\cite{chime9}. Figure~\ref{fig:grid} shows the first gridding observation of FRB~20190303A on 4 Jan 2021 and 5 Jan 2021. One burst was detected in Beam M13 of the N3 pointing with position of (RA = 13$^h$51$^m$58$^s$, Dec = 48$^\circ$07$'$20$''$) (J2000).

We then carried out the third and fourth session on 19 Jan 2021 and 20 Jan 2021 to do a second 2-hour gridding observation. 98.304\,$\upmu$s time resolution was used. The central beam of the receiver was initially placed on the position of the first detection. One burst was detected in Beam M01 of the N1 pointing, i.e. the position of the first detection. 

Because we did not detect any burst signal in other beams, we take (RA = 13$^h$51$^m$58$^s$, Dec = +48$^\circ$07$'$20$''$) (J2000) as the probable FRB position and carried out another two observations using only the central beam placed on the probable FRB position. 49.152\,$\upmu$s time resolution was used. The fifth session on 3 Feb 2021 lasted for two hours and no bursts were detected. The sixth session on 14 Feb 2021 lasted for one hour and two bursts were detected.   

The tracking data was searched using the \textsc{fast\_miner} pipeline\cite{2021ApJ...909L...8N}. The data stream from each beam was processed using \textsc{heimdall} \cite{2012MNRAS.422..379B}, which  dedisperses the data incoherently with DM ranging between 20 and 2000  $\rm pc\ cm^{-3}$.
We kept the candidates that show less than 4 adjacent beams and S/N greater than 7 for further identification from waterfall plots. From the waterfalls we identified four bursts reported above. 

\subsection{FRB~20190417A}
The observations of FRB~20190417A were carried out using the FAST telescope mounted with
the 19-beam receiver. Observations were carried out in two sessions. We carried out the first and second session on 30 Jul 2020 and 17 Aug 2020 to do a 1-hour observation using all beams of the 19-beam receiver. 49.152 and 98.304 \,$\upmu$s time resolution were used. The central beam of the receiver was initially placed on the previously reported position (RA = 19h39min, Dec = +59$^\circ$24$'$) (J2000)\cite{chime9}. Two bursts were detected in Beam M07 with position of (RA = 19$^h$39$^m$22$^s$, Dec = +59$^\circ$18$'$58$''$) (J2000). The central beam of the receiver was placed on the position of the first detection. 23 bursts were detected in Beam M01, i.e. the position of the first detection.     

The tracking data was searched by dedicated and blind search during the observational campaign. We performed 14 box-car pulse width match-filter grids scheduled in logarithmic space from 0.1 to 30 ms using \textsc{presto}. A zero-DM matched filter was applied to mitigate radio frequency interference (RFI) in the blind search. The data were de-dispersed at DMs ranging from 800 to 1800\,pc~cm$^{-3}$ with the step size of 1\,pc~cm$^{-3}$. All of the possible pulse candidates of S/N$>$6 were plotted, then be confirmed pulse by pulse with dispersive signature to remove the narrow-band RFI.

\subsection{FRB~20201124A}

FRB~20201124A was observed with the 800 MHz feed ( 720--920 MHz) of the GBT's prime focus receiver and the VEGAS digital backend.  The data were coherently dedispersed at a DM of 413.52 pc~cm$^{-3}$ and recorded in the \textsc{psrfits} standard format.  Full polarization self and cross products were recorded every 81.92 $\upmu$s with 195.3125 kHz-wide channels (i.e., 1024 frequency channels). Data were calibrated using the same procedure as we used for FRB~20190520B, except we used FIRST J141341.6+15093 (aka J1413+1509) as the calibration source.  

We searched for bursts using \textsc{presto} in a similar way as with FRB~20190520B, with the main differences being that we retained the full time and frequency resolution and searched DMs ranging from 200--600 pc~cm$^{-3}$ with a step size of 1 pc~cm$^{-3}$. 9 bursts were detected. 

FRB~20201124A was also observed with the FAST  telescope between 1.0 and 1.4 GHz. The data streams were processed with the ROACH2 signal processor and recorded as 8 bit-sampled search mode \textsc{psrfits} files with 4096 frequency channels and 49.152\,$\upmu$s sampling time. Using \textsc{presto}, we searched DMs ranging from 200--600 pc~cm$^{-3}$ with a step size of 1 pc~cm$^{-3}$, resulting in 11 bursts.

\section{Polarization and Faraday rotation}
The bursts were dedispersed using previously published dispersion measures, listed in Table~S2. Polarization calibration was performed with the \textsc{psrchive} software package \cite{2004PASA...21..302H} using the single-axis model. This calibration strategy uses a noise diode to correct for the differential gain and phase between the two polarization channels. This calibration scheme does not correct for leakage. At FAST, the leakage term is better than -46 dB within the full width at half maximum region of the central beam as measured during the FAST engineering phase \cite{FAST19Beam}, which corresponds to systematic errors less than 0.5\%. The polarization properties of bright pulsars are consistent between FAST and the Parkes Pulsar Timing Array within 0.5\% \cite{dai15}. The cross-polarization leakage of the GBT C-Band receiver when recording linear polarization is $< -32$ dB over our observed bandwidth, and over most of the band it is $< -37$ dB, which corresponds to systematic errors of about 1.4\%.  To excise RFI, we used the \textsc{psrchive} software package to median filter each burst in the frequency domain and mitigated RFI of each burst manually. 

\begin{figure}[!htp]
\centering

\subfigure[FRB 20190303A]{
    \includegraphics[width=0.45\textwidth]{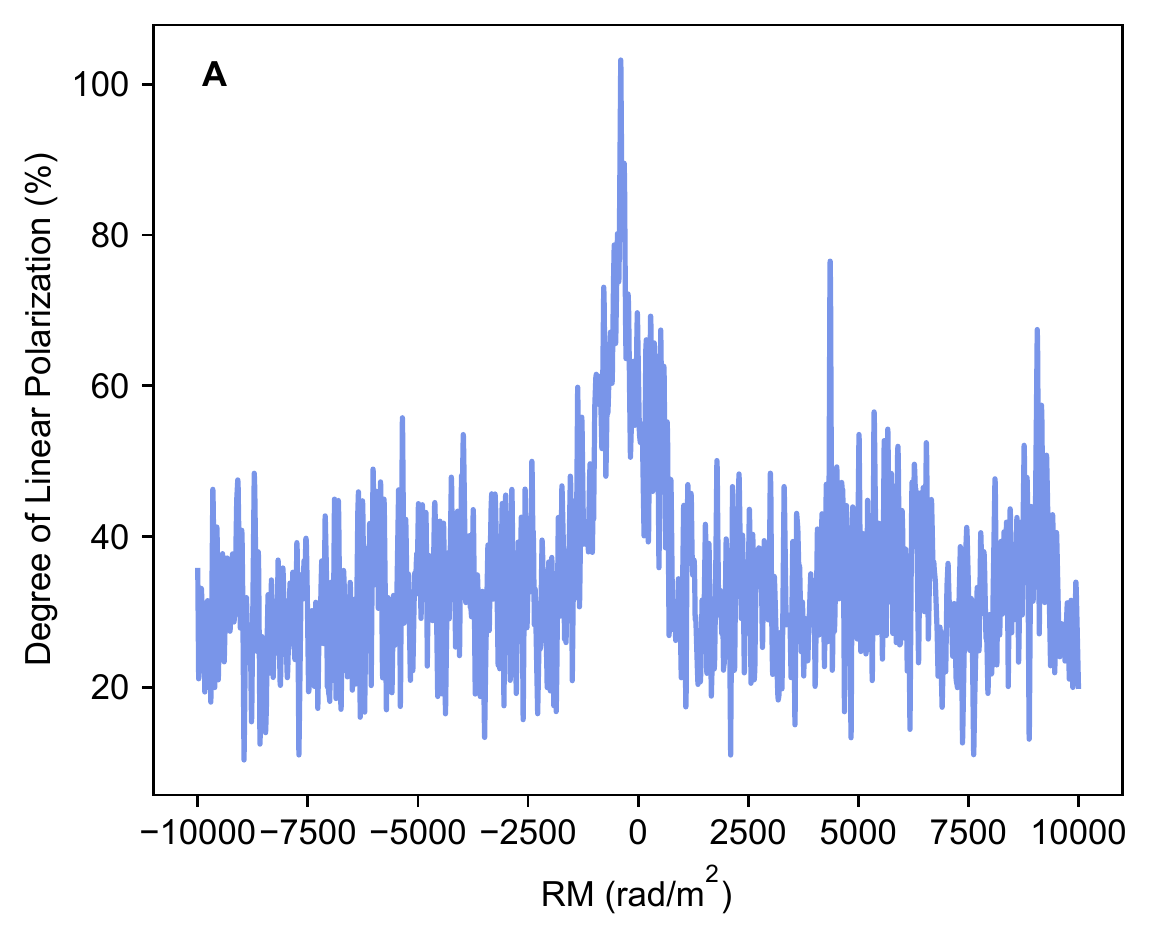}
}
\subfigure[FRB 20190417A]{
    \includegraphics[width=0.45\textwidth]{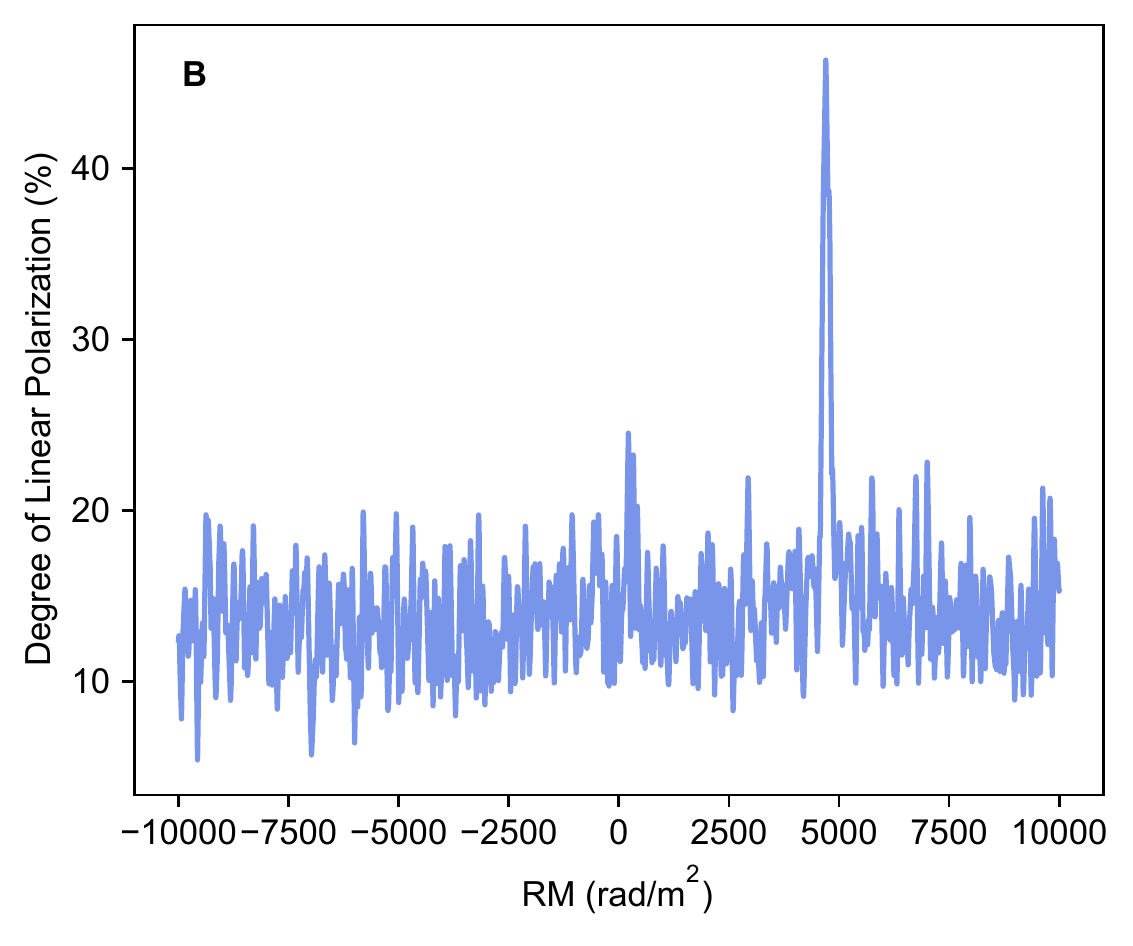}
}\\
\subfigure[FRB 20190520B]{
    \includegraphics[width=0.45\textwidth]{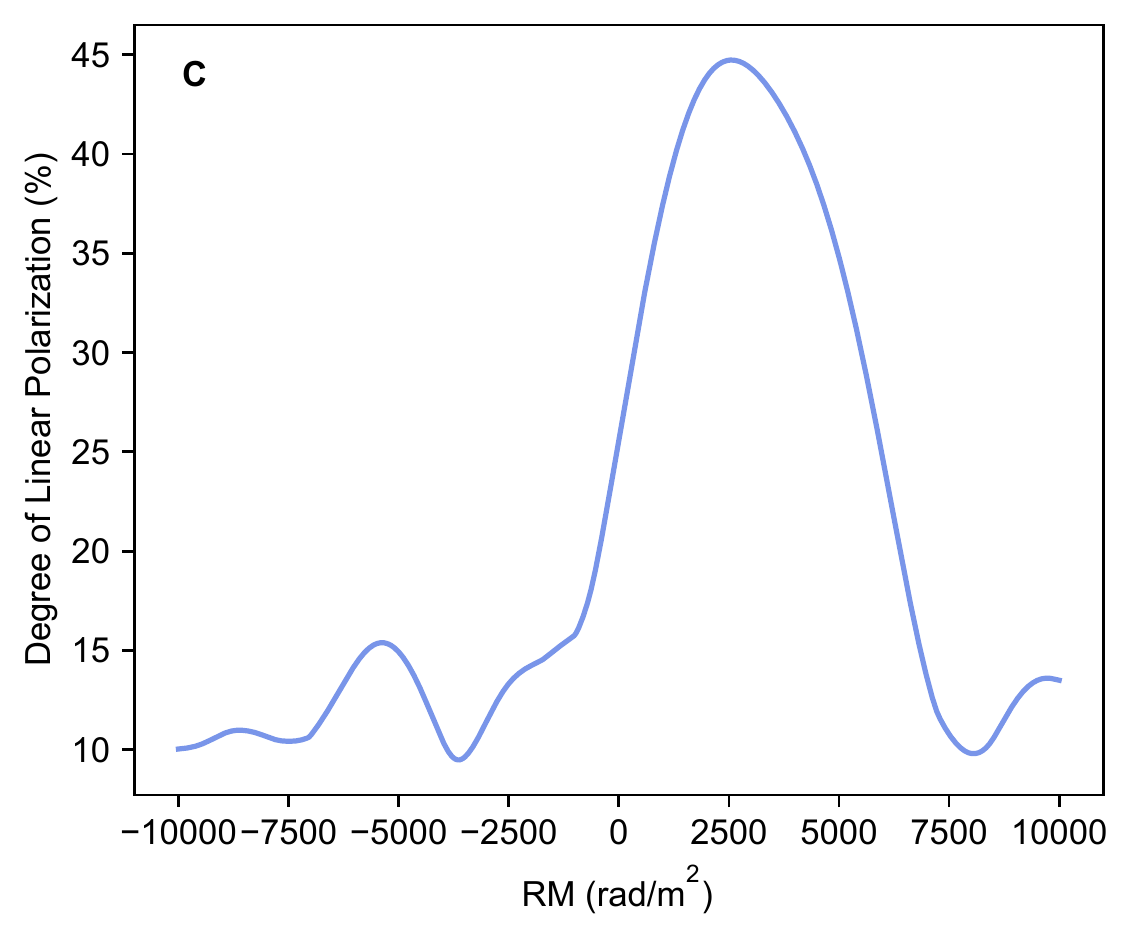}
}
\subfigure[FRB 20201124A]{
    \includegraphics[width=0.45\textwidth]{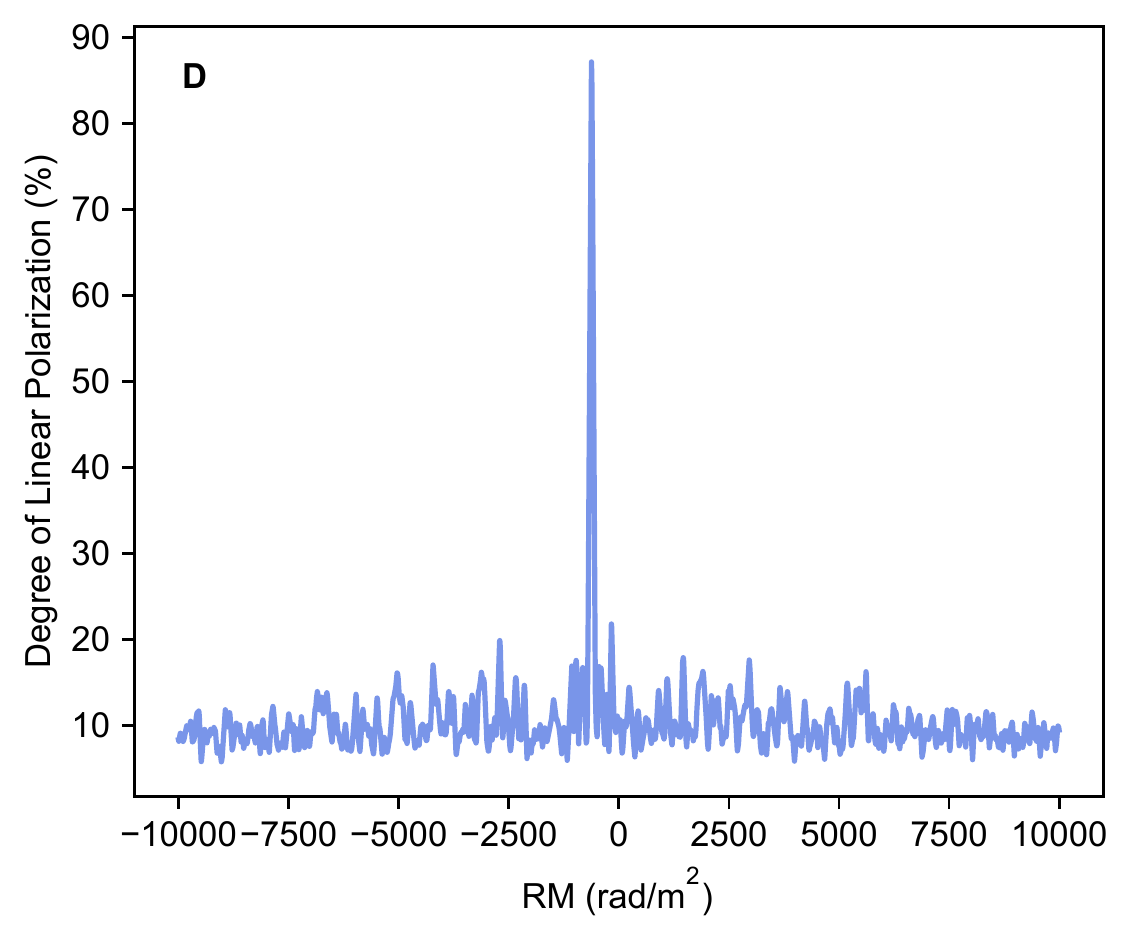}
}

\caption{\textbf{RM search with RM-synthesis.} Example results of RM-synthesis for each FRB with RM detection. The blues lines represent linear polarization fraction of the bursts as 
a function of rotation measure.}
\label{fig:rm-fdf}
\end{figure}

\begin{figure}[!htp]
\centering

\subfigure[FRB 20190303A]{
    \includegraphics[width=0.45\textwidth]{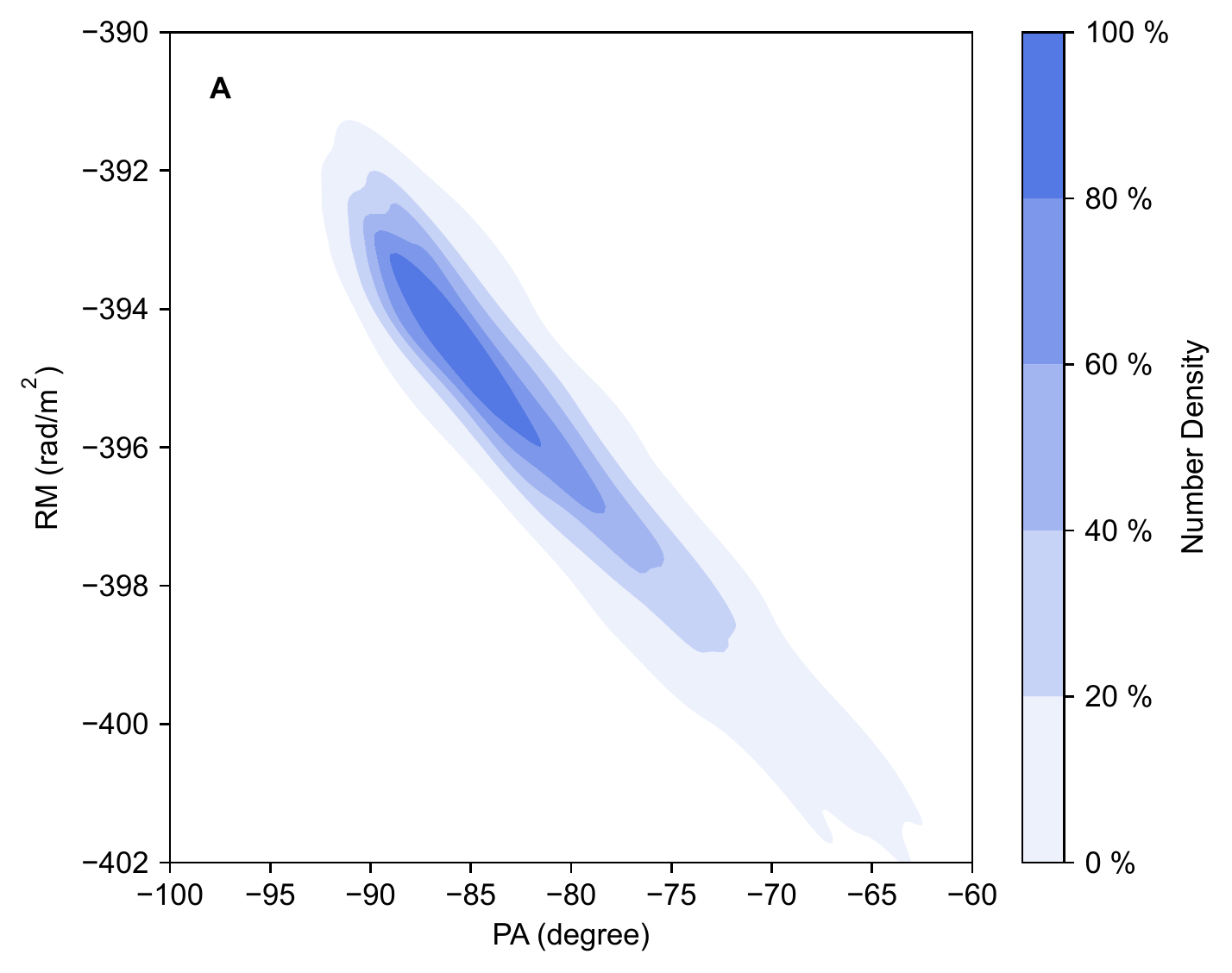}
}
\subfigure[FRB 20190417A]{
    \includegraphics[width=0.45\textwidth]{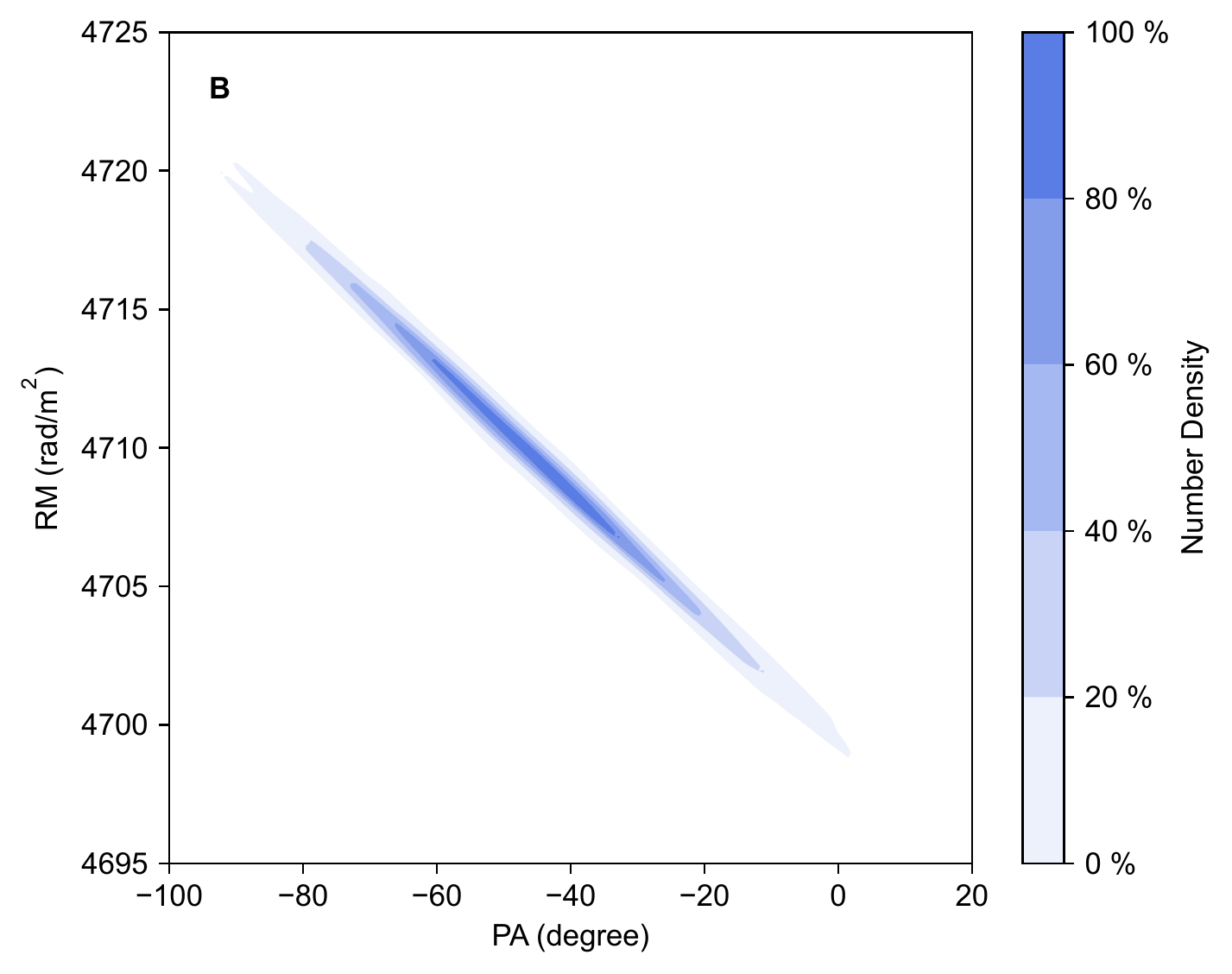}
}\\
\subfigure[FRB 20190520B]{
    \includegraphics[width=0.45\textwidth]{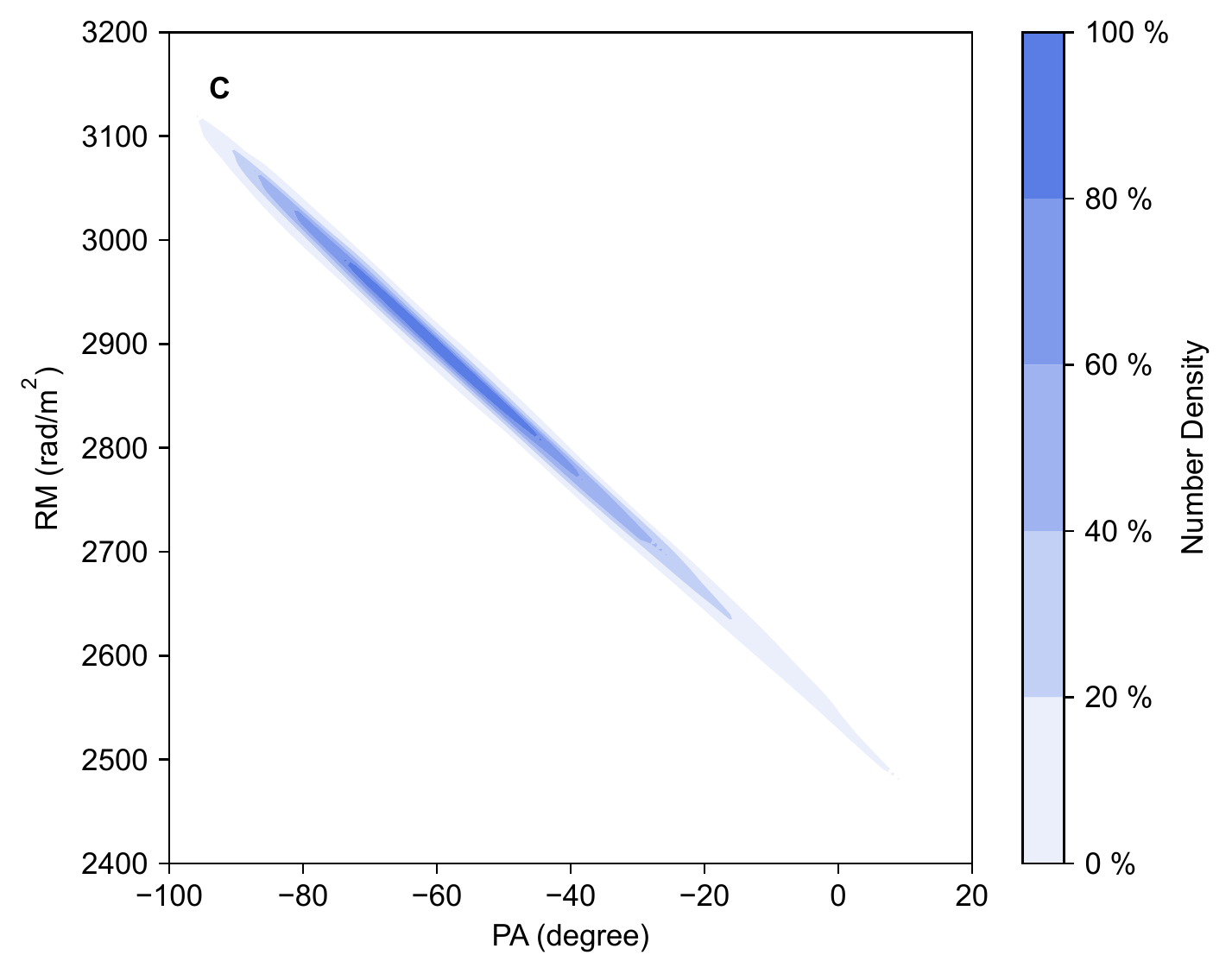}
}
\subfigure[FRB 20201124A]{
    \includegraphics[width=0.45\textwidth]{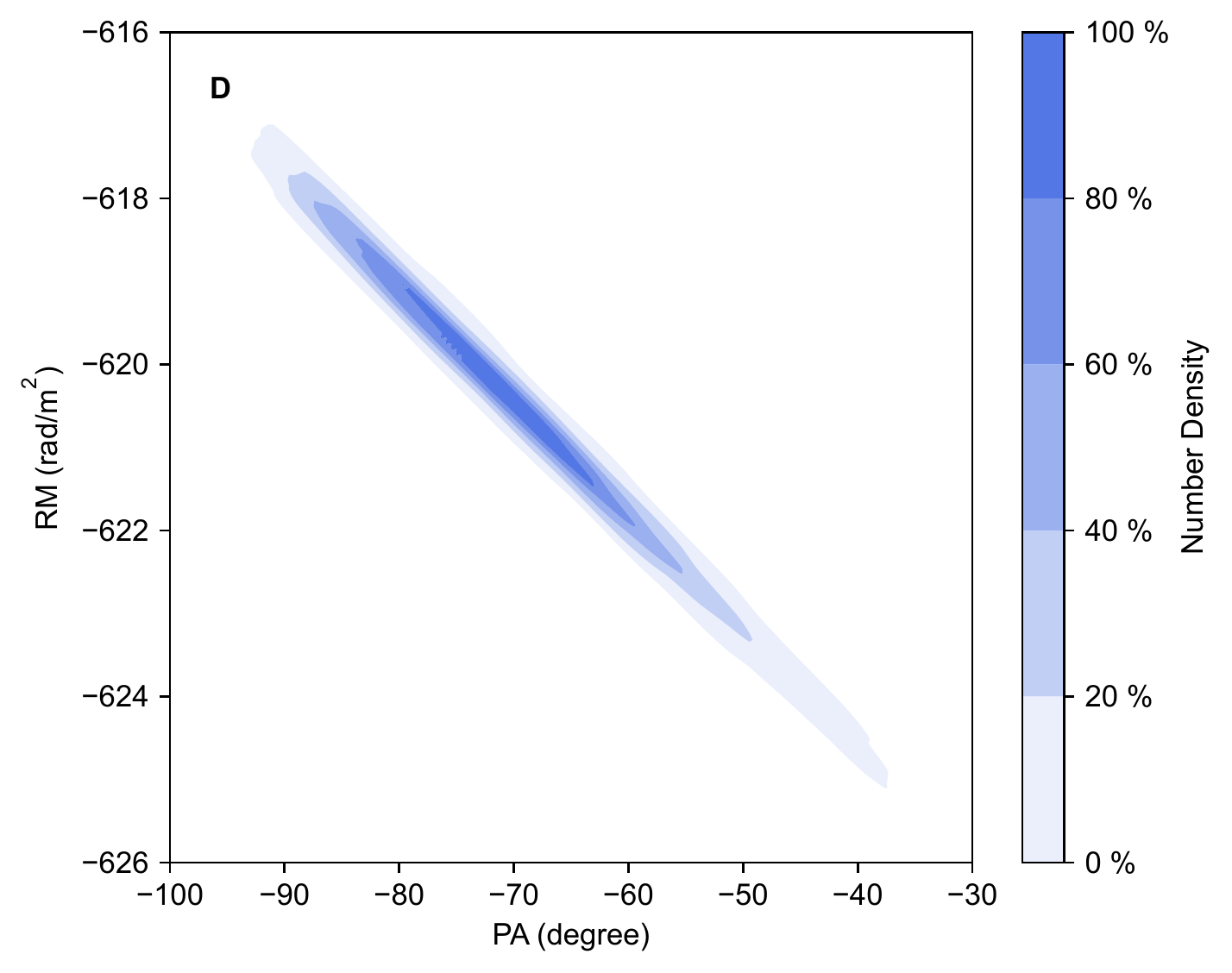}
}

\caption{\textbf{RM search with Stokes QU-fitting.} Example results of Stokes QU-fitting for the same bursts shown in Figure~\ref{fig:rm-fdf}. Each panel shows the two dimensional posterior probability distributions of the RM and PA for FRB~20190303A (Panel A), FRB~20190417A (Panel B), FRB~20190520B (Panel C) and FRB~20201124A (Panel B). The selection of contour levels is displayed in the colour bar.}
\label{fig:rm-mcmc}
\end{figure}

The RM is defined as
\begin{equation} \label{eq:rm}
    \text{RM} \equiv 0.81 \int_{d}^{0} \frac{B_\parallel(l) n_e(l)}{(1+z)^2} dl,
\end{equation}
where $l$ is the line-of-sight position; $B_\parallel$ is the line-of-sight magnetic field strength in microgausss; $n_e$ is the electron density; $z$ is the redshift of the source; and $d$ is the distance to the source. We report the observed RM and do not correct it for redshift,. We searched for an RM detection using the methods of Stokes QU-fitting \cite{2012MNRAS.421.3300O} and RM-synthesis \cite{1966MNRAS.133...67B, 2005A&A...441.1217B}. The RM values of FRB~20190520B, FRB~20190303A, FRB~20190417A, and FRB~20201124A are listed in Table~\ref{tab:burst}. Examples of the results from RM-synthesis for each FRB with RM detection are shown in Figure~\ref{fig:rm-fdf} and for Stokes QU-fitting in Figure~\ref{fig:rm-mcmc}. We find consistent values with both methods (Table S3).

\begin{figure}[!htp]
\centering
\subfigure[FRB 20121102A]{\includegraphics[width=0.95\textwidth]{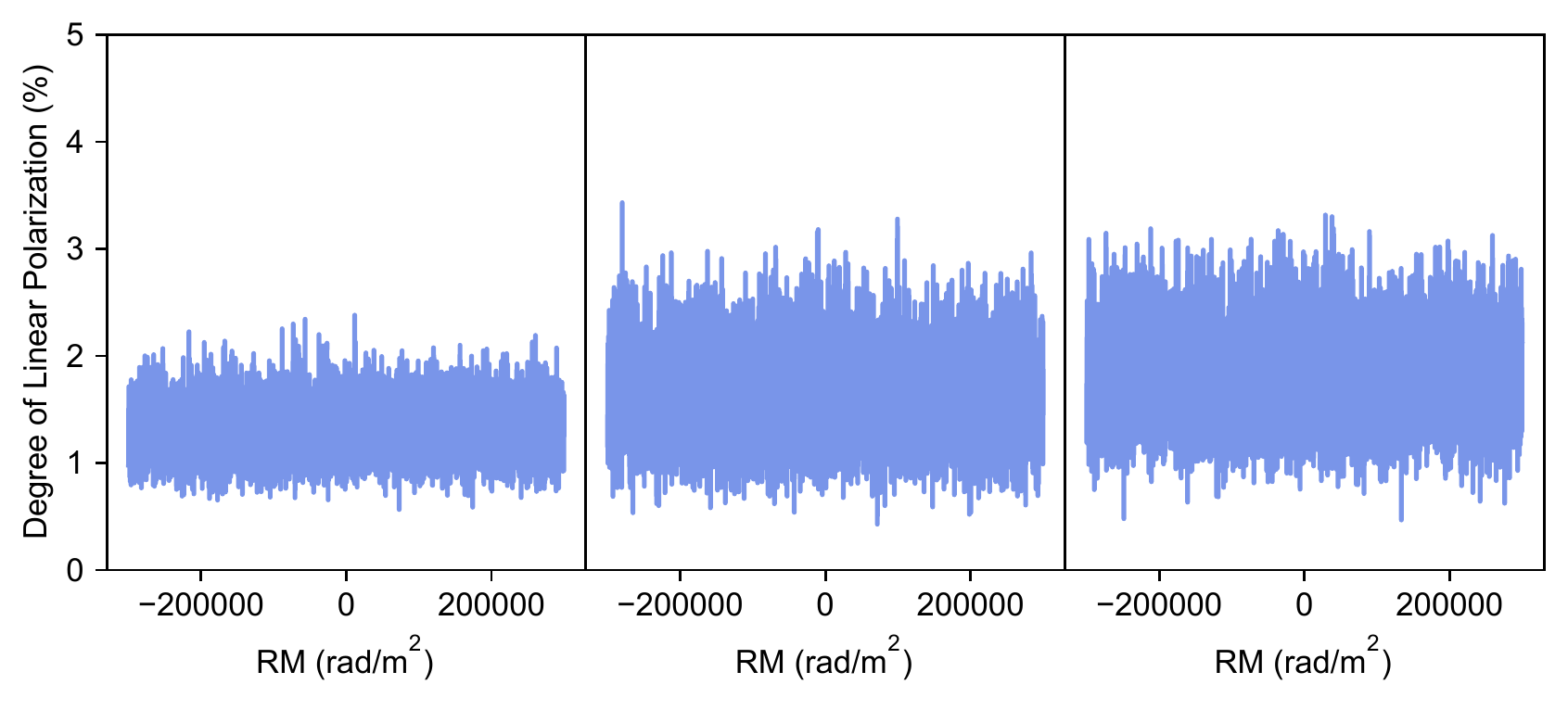}
}\\
\subfigure[FRB 20190520B]{\includegraphics[width=0.95\textwidth]{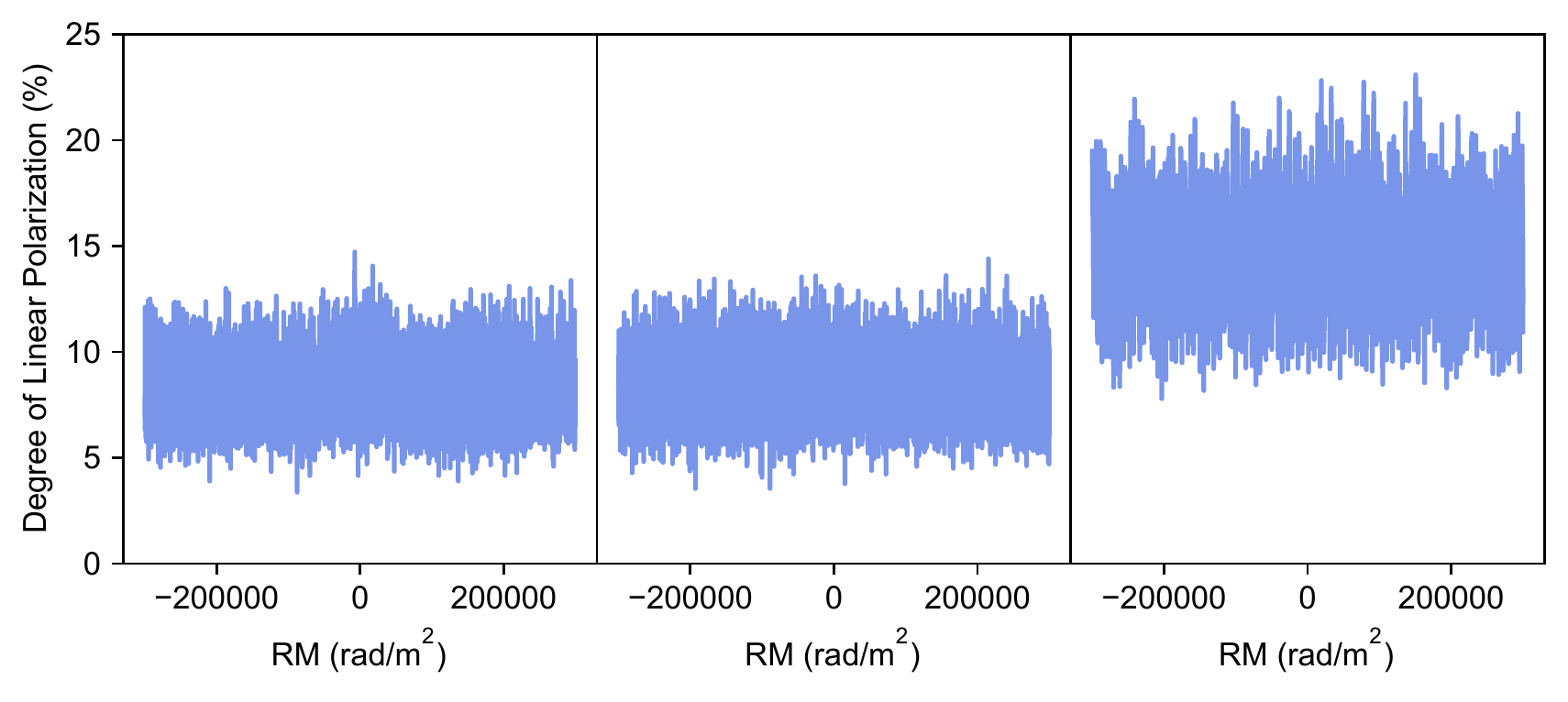}
}
\caption{\textbf{RM search for FRBs~20121102A and 20190520B.} RM search for the three brightest bursts of FRB~20121102A and FRB~20190520B at 1.0-1.5\,GHz with FAST. No peak was found in the Faraday spectrum.}
\label{fig:rm_190502}
\end{figure}

For FRB~20121102A, we obtained polarization measurements resulting in non-detection of linear polarization at 1.0-1.5\,GHz with FAST. We searched for the RM from $-3.0\times10^5$ to $3.0\times10^5$\,$\mathrm{rad\,m^{-2}}$ for all bursts of FRB~20121102A at 1.0-1.5\,GHz with FAST and we show the RM search for the three brightest bursts in Figure~\ref{fig:rm_190502}. No peak was found in the Faraday spectrum and we place an upper limit of 6\% on the degree of linear polarization for FRB~20121102A at 1.25\,GHz. 

For FRB~20190520B, we obtained polarization measurements resulting in non-detection of linear polarization at 1.0-1.5\,GHz with FAST. We searched for the RM from $-3.0\times10^5$ to $3.0\times10^5$\,$\mathrm{rad\,m^{-2}}$ for three brightest bursts of FRB~20190520B at 1.0-1.5\,GHz with FAST and the result is shown in Figure~\ref{fig:rm_190502}. No peak was found in the Faraday spectrum and we place an upper limit of 20\% on the degree of linear polarization for FRB~20190520B at 1.25\,GHz. $f_{\rm depol}$ = 0.2 when $\mathrm{RM} = 10^5\,\mathrm{rad/m}^2$, therefore the non-detection of linear polarization of FRB~20121102A and 20190520B at 1.0-1.5\,GHz is not caused by intra-channel depolarization.   

Finally, we calculated the degrees of linear polarization and circular polarization for bursts with RM detection. We first derotated the linear polarization with the measured RM. The representative RM-corrected polarization profiles of FRB~20190303A, FRB~20190520B, FRB~20190417A, and FRB~20201124A are shown in Figure 1. The measured linear polarization is overestimated in the presence of noise. Therefore we use the frequency-averaged, de-biased total linear polarization \cite{2001ApJ...553..341E} :
\begin{equation} \label{eq:L_de-bias}
    L_{{\mathrm{de\mbox{-}bias}}} =
    \begin{cases}
      \sigma_I \sqrt{\left(\frac{L_{i}}{\sigma_I}\right)^2 - 1} & \text{if $\frac{L_{i}}{\sigma_I} > 1.57$} \\
      0 & \text{otherwise} ,
    \end{cases}
\end{equation}
where $\sigma_I$ is the Stokes I off-pulse standard deviation and $L_i$ is the measured frequency-averaged linear polarization of time sample $i$.
We defined the degree of linear polarization 
as ($\Sigma_{i} L_{\mathrm{de\mbox{-}bias},i}$)/($\Sigma_{i}I_i$)
and that of circular polarization as ($\Sigma_{i} V_i$)/($\Sigma_{i}I_i$), where the summation is over the bursts and $V_i$ is the measured frequency-averaged circular polarization of time sample $i$. 
Defining $I = \Sigma_{i}I_i$, $L = \Sigma_{i} L_{\mathrm{de\mbox{-}bias},i}$ and $V = \Sigma_{i}V_i$, uncertainties on the linear polarization fraction and circular polarization fraction are calculated as:
\begin{equation} \label{eq:uncertainty}
    \sigma_{\rho/I} = \frac{\sqrt{N+N\frac{\rho^2}{I^2}}}{I}\sigma_{I},
\end{equation}
where $N$ is the number of time samples of the burst, and $\rho = L,V$ for linear and circular polarization fraction, respectively.
The degrees of linear polarization and circular polarization are listed in Table~\ref{tab:burst}.

\section{RM budget and $\sigma_{\mathrm{RM}}$}

The observed total rotation measure $\mathrm{RM_{tot}}$ is a combination of multiple contributions:
\begin{equation}
\label{eq:rm_com}
\mathrm{RM_{tot}}=\mathrm{RM_{iono}}+\mathrm{RM_{Gal}}+\mathrm{RM_{IGM}}+\mathrm{RM_{host}}+\mathrm{RM_{source}}, 
\end{equation}
where $\mathrm{RM_{iono}}$ is the RM due to the Earth’s ionosphere, $\mathrm{RM_{Gal}}$ is the Galactic foreground, $\mathrm{RM_{IGM}}$is the contribution from the intergalactic medium
(IGM), $\mathrm{RM_{host}}$ is the RM in a host galaxy and $\mathrm{RM_{source}}$ is the intrinsic component from magnetised plasma associated with the progenitor source and its immediate environment. The observed $\mathrm{RM_{host}}$ and $\mathrm{RM_{source}}$ are smaller by $(1+z)^2$ compared with their rest-frame values (Eq.~\ref{eq:rm}), and our RM budget assumes that the observed values are the same as the rest-frame values to simplify the analysis. $\mathrm{RM_{iono}}$ does not exceed a few $\mathrm{rad/m}^2$\cite{2013A&A...552A..58S}. $\mathrm{RM_{IGM}}$ is typically smaller than 20\,$\mathrm{rad/m}^2$ \cite{rm_igm} and $\mathrm{RM_{Gal}}$ is typically smaller than 100\,$\mathrm{rad/m}^2$ \cite{rmsampling}. $\mathrm{RM_{host}}$ can be reasonably assumed to be smaller than $\mathrm{RM_{Gal}}$. Thus we assume $\mathrm{RM_{tot}}=\mathrm{RM_{source}}$ for the repeaters in our sample. 

For each repeater which has bursts with degree of linear polarization less than 90\%, we determine $\sigma_{\mathrm{RM}}$ by fitting the data with the model in Eq.~2 assuming each burst has 100\% intrinsic linear polarization. For FRB~20201124A, an unweighted least squares fit was used because the FAST measurements have much smaller uncertainties than the GBT data. For other FRBs, an inverse-variance weighted least squares fit was used. For $\lambda$ in Eq.~2, we calculated the central frequency of each burst weighted by signal-to-noise ratio and converted the weighted frequency to the equivalent wavelength $\lambda$. The resulting $\sigma_{\mathrm{RM}}$ values are listed in Table~\ref{tab:frb_property}.
The inferred $\sigma_{\mathrm{RM}}$ is in the observer's frame. If the source were at a high redshift $z$, then the true $\sigma_{\mathrm{RM}}$ is higher by a factor of $(1+z)^2$, which is the same scaling as RM. For instance, if FRB~20190417A, which has a large DM of 1378\,pc~cm$^{-3}$ \cite{chime9}, is at z = 1, then the true $\sigma_{\mathrm{RM}}$ is a factor of 4 larger, making this source similar to FRB~20121102A.

\paragraph{\bf\LARGE Supplementary Text}
\section{RM scatter and temporal scattering caused by a magnetised plasma screen}

\begin{figure}[!htp]
\centering
\includegraphics[width = 1.0\linewidth, trim = 100 200 100 200, clip]{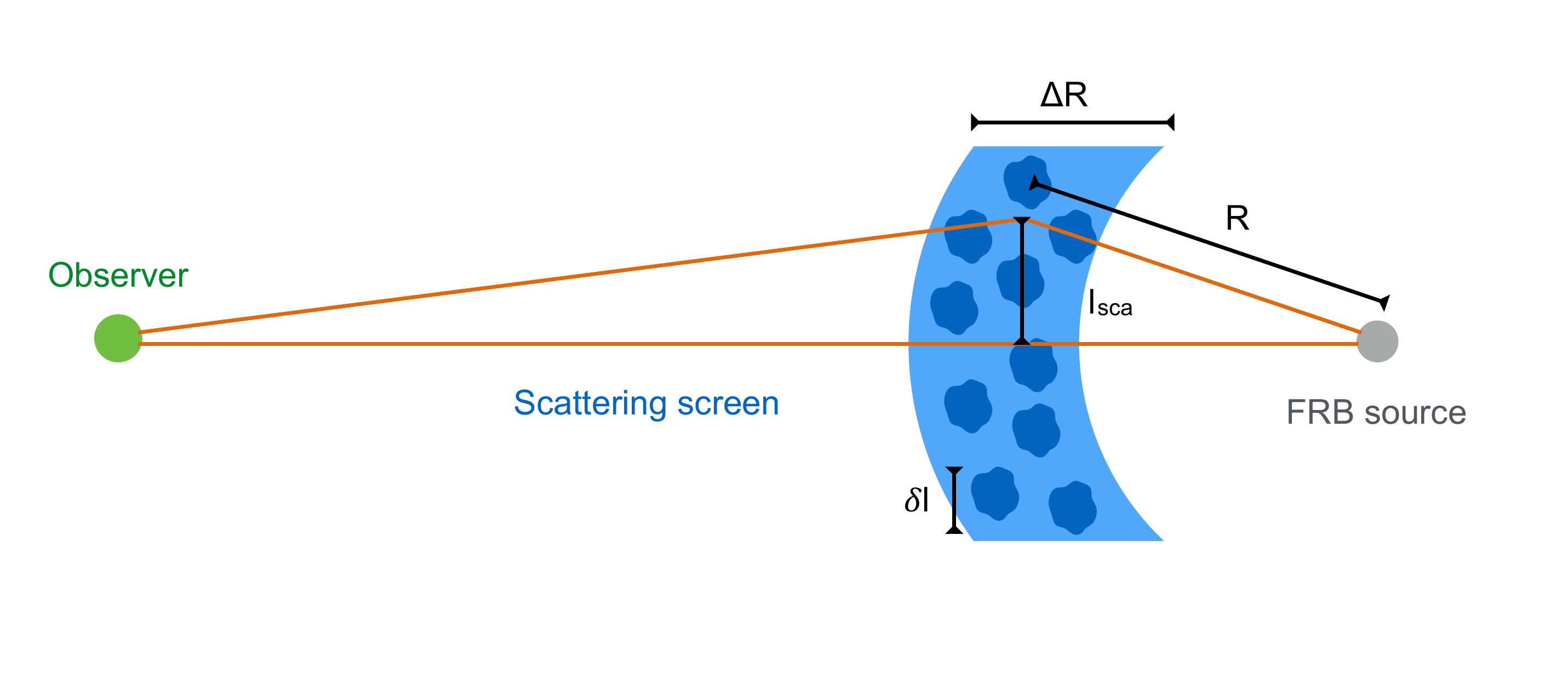}
\caption{\textbf{A Schematic of temporal scattering and RM scatter induced by multi-path propagation.} Schematic configuration of FRB propagating in the magneto-ionic environments as the plasma screen around the FRB source.}
\label{fig:rmscatter}
\end{figure}

In this section, we discuss scintillation, temporal scattering, and $\sigma_{\rm RM}$ from a magneto-ionic inhomogeneous plasma screen near the repeater source, as shown schematically in Figure \ref{fig:rmscatter} (see also Ref.~\cite{2022MNRAS.510.4654B} for a more detailed treatment). 
We consider that the plasma screen has a thickness $\Delta R$ and a radius $R$, respectively, and assume $\Delta R\sim R$. The electron density fluctuation is $\delta n_e$ in the scale $\delta l$. $\delta l/R<\theta_{\rm s}$ is required for scintillation and temporal scattering, where $\theta_{\rm s}$ is the refraction angle from Eq.(\ref{thetas}).
When a radio burst propagates in the plasma screen, the variation in phase velocity is approximately $\delta v_{\rm pha}\simeq c\delta(\omega_p^2)/2\omega^2=2\pi e^2c\delta n_e/m_e\omega^2$ for $\omega\gg\omega_B$, where $m_e$ is the electron rest mass, $\omega_p=(4\pi e^2 n_e/m_e)^{1/2}$ is the plasma frequency, $\omega_B=eB/m_ec$ is the electron cyclotron frequency, and $\omega$ is the wave angular frequency. 
The wavefront is advanced or retarded by $\sim(\delta v_{\rm pha}/c)\delta l$ relative to propagation in a homogeneous medium, after the wavefront traverses a single clump of lengthscale $\delta l$. 
After traveling through the plasma screen with thickness $\Delta R$, the wavefront has crossed $\Delta R/\delta l$ clumps, and the advance or retardation of the wavefront is $\delta x\sim(\Delta R/\delta l)^{1/2}(\delta v_{\rm pha}/c)\delta l$.
The wavefront surface will be tilted by an angle of $\theta_{\rm s}\simeq\delta x/\delta l$ due to multiple refraction by clumps, and the characteristic value of the refraction angle is
\be
\theta_{\rm s}&\simeq&\left(\frac{\Delta R}{\delta l}\right)^{1/2}\frac{e^2}{2\pi m_e\nu^2}\delta n_e=7.1\times10^{-6}~{\rm rad}\left(\frac{\Delta R}{1~{\rm pc}}\right)^{1/2}\nonumber\\
&\times&\left(\frac{\delta l}{10^{12}~{\rm cm}}\right)^{-1/2}\left(\frac{\delta n_e}{100~{\rm cm^{-3}}}\right)\left(\frac{\nu}{1~{\rm GHz}}\right)^{-2},\label{thetas}
\ee
where $\nu=\omega/2\pi$ is the wave frequency.
From Figure \ref{fig:rmscatter}, the path-length difference between two rays is $\Delta s=R(1-\cos\theta_{\rm s})\simeq R\theta_{\rm s}^2/2$ for $\theta_{\rm s}\ll1$. Thus, the temporal scattering time can be estimated as 
\be
\tau_{\rm sca}&\simeq&\frac{R\theta_{\rm s}^2}{2c}=2.6~{\rm ms}\left(\frac{R}{1~{\rm pc}}\right)^{2}\nonumber\\
&\times&\left(\frac{\delta l}{10^{12}~{\rm cm}}\right)^{-1}\left(\frac{\delta n_e}{100~{\rm cm^{-3}}}\right)^{2}\left(\frac{\nu}{1~{\rm GHz}}\right)^{-4}\label{tau}.
\ee
Here, the assumption of $\Delta R\sim R$ is used.
For a certain plasma screen, scintillation and temporal scattering occur together and have a relationship of $\Delta \nu_{\rm sci}=1/(2\pi\tau_{\rm sca})$, where $\Delta \nu_{\rm sci}$ is the scintillation bandwidth.
The corresponding scintillation bandwidth is
\be
\Delta\nu_{\rm sci}&\simeq&\frac{1}{2\pi \tau_{\rm sca}}=62~{\rm Hz}\left(\frac{R}{1~{\rm pc}}\right)^{-2}\nonumber\\
&\times&\left(\frac{\delta l}{10^{12}~{\rm cm}}\right)\left(\frac{\delta n_e}{100~{\rm cm^{-3}}}\right)^{-2}\left(\frac{\nu}{1~{\rm GHz}}\right)^{4},
\ee
This value is much smaller than the observed scintillation bandwidth of $\Delta\nu_{\rm sci}\sim1{\rm MHz}$. This is consistent with the observed scintillation being more likely contributed by the interstellar medium within the Milky Way.

Next, we discuss the effect of RM scatter. We assume that the magnetic field also has a typical lengthscale $\delta l$. 
The differences of multi-path RMs is large enough to cause depolarization when a radio wave propagates in the magneto-ionic inhomogeneous environment.
The decorrelation lengthscale of the fluctuation of Faraday rotation angle could be much larger than that of the fluctuation of wave phase in the plasma screen, because the evolution of the rotation angle is much slower than that of the wave phase for $\omega\gg\omega_B$. Because the large-scale turbulence dominates the Faraday rotation, the decorrelation lengthscale of Faraday rotation angle is approximately the transverse lengthscale of the visible part of the plasma screen, i.e.,
\be
l_{\rm sca}&=&\theta_{\rm s}R=2.2\times10^{13}~{\rm cm}\left(\frac{R}{1~{\rm pc}}\right)^{3/2}\nonumber\\
&\times&\left(\frac{\delta l}{10^{12}~{\rm cm}}\right)^{-1/2}\left(\frac{\delta n_e}{100~{\rm cm^{-3}}}\right)\left(\frac{\nu}{1~{\rm GHz}}\right)^{-2}.\label{lsca}
\ee
According to the definition of rotation measure, i.e., $\phi\equiv\lambda^2{\rm RM}$, where $\phi$ is the rotation angle, $\lambda$ is the wavelength, and RM is the rotation measure, the total RM scatter contributed by the plasma screen could be calculated by 
\be
\sigma_{\rm RM}&\simeq&\sqrt{\frac{\Delta R}{l_{\rm sca}}}|{\rm RM_{sca}}|=1.4~{\rm rad~m^{-2}}\left(\frac{R}{1~{\rm pc}}\right)^{1/2}\nonumber\\
&\times&\left(\frac{\delta (n_e B_\parallel)}{10^3~{\rm cm^{-3}~\upmu G}}\right)\left(\frac{l_{\rm sca}}{10^{13}~{\rm cm}}\right)^{1/2}\label{sigma},
\ee
where the RM contribution in a turbulence region with scale $l_{\rm sca}$ is given by
\be
{\rm RM_{sca}}=\frac{e^3}{2\pi m_e^2c^4} \delta(B_\parallel n_e) l_{\rm sca}.
\ee
The requirements of $\delta(n_eB_\parallel)\sim10^3~{\rm cm^{-3}\upmu G}$ on the scale $l_{\rm sca}\sim10^{13}~{\rm cm}$ and $R\sim1~{\rm pc}$ suggest that the RM scatter and temporal scattering originate from a radio burst propagating in an inhomogeneous magneto-ionic  environment near the repeating source, which may be related to a supernova remnant (SNR) or a wind nebula (PWN). For example, the observations of SNRs show that they have a magnetic field strength $B\sim{\rm a~few}~{\rm \upmu G}$ to ${\rm a~few}~{\rm m G}$ \cite{2012SSRv..166..231R}. For an SNR with the ejecta mass $M_{\rm ej}$, the electron number density at $R$ could be estimated as
\be
n_e\simeq\frac{M_{\rm ej}}{m_p(4\pi/3)R^3}=98~{\rm cm^{-3}}\left(\frac{M_{\rm ej}}{10M_\odot}\right)\left(\frac{R}{1~{\rm pc}}\right)^{-3}
\ee
where $m_p$ is the proton mass, and the ejecta is assumed to be dominated by hydrogen here. Thus, $\delta(n_eB_\parallel)\sim10^3~{\rm cm^{-3}\upmu G}$ at $R\sim1~{\rm pc}$ is attainable for an SNR.
On the other hand, according to Eq.(\ref{tau}) and Eq.(\ref{lsca}), $l_{\rm sca}\sim10^{13}~{\rm cm}$ is consistent with the observed temporal scattering time.
In conclusion, the required parameters in Eq.(\ref{sigma}) are attainable for the SNR scenario (see Refs.\cite{2017ApJ...847...22Y,2018ApJ...861..150P} for more detailed theoretical treatments of SNR properties). 
At last, the DM contribution from the plasma screen is given by
\be
{\rm DM}\simeq n_e R=100~{\rm pc~cm^{-3}}\left(\frac{n_e}{100~{\rm cm^{-3}}}\right)\left(\frac{R}{1~{\rm pc}}\right).
\ee
This explains the large $\rm DM_{host}$ values observed in some of these repeating FRBs, especially FRB 20121102 \cite{2016Natur.531..202S} and FRB 20190520 \cite{niu21}.
\clearpage

\clearpage

\begin{center}
\begin{ThreePartTable}
\begin{TableNotes}
  \item [*] {We use value in Table 1 in Ref. \cite{luo2020} obtained by RM synthesis for FRB~20180301A.}
  \item [$\dagger$] {This is a lower bound due to possible leakage of signal of Stokes U into Stokes V.\cite{chime9}}
  \item [$\ddagger$] {Value of sub-burst 1. Sub-burst 2 and sub-burst 3 are $94\pm2$\% and $98\pm4$\% linear polarized.}
  \item [$\S$] {Value of sub-burst 1. Sub-burst 2 and sub-burst 3 are $1\pm2$\% and $1\pm3$\% circular polarized.}
  \item [$\rVert$] {RM $>$118000\,rad~m$^{-2}$ would cause depolarization.}
  \item [\P] {The linear polarization could be depolarized and the RM could be large.}
  \item [\#] {Value of the first pulse.}
  \item [**] {The degree of circular polarization shows a variation across the pulse, and we choose the largest absolute value of -34.}
  \item [$\dagger\dagger$] {Value of sub-burst 2.}
\end{TableNotes}
\begin{scriptsize}
\begin{longtable}{cccccccccc}
  \rmfamily\\
  \caption{\textbf{Polarization properties of published FRBs.} For multiple bursts, if the bursts have nearly 100\% linear polarization, we just include the brightest burst, otherwise we include each of the burst. Column (1): name of the FRB;  Col.(2): Repetition; Col.(3): the telescope of the observation; Col.(4): frequency of the burst weighted by signal to noise ratio for repeaters. For non-repeaters, it is the representative frequency of the receiver; Col.(5): channel frequency width; Col.(6): fractional reduction in the linear polarization amplitude; Col.(7): rotation measure; Col.(8): degree of linear polarization; Col.(9): degree of circular polarization; Col.(10): references. `-' represents not reported in the reference.}\label{tab:data}\\
  \toprule
  Name & Repeater & Telescope & Frequency & $\Delta\nu$ & $f_{\mathrm{depol}}$ & RM &  \%~Linear & \%~Circular & Ref \\
                   &      &           &    (MHz)    &    (MHz)      &             & (rad~m$^{-2}$)  &       \\
  \midrule
  \endfirsthead

  \toprule
  \endhead
  
  \bottomrule
  \endfoot

  \bottomrule
  \insertTableNotes
  \endlastfoot

 FRB~20121102A & yes  &AO   & 4600 & 1.56 & $7.1\times10^{-3}$  & $71525\pm3$        & $95.2\pm0.4$   & -      & \cite{2021ApJ...908L..10H}\\
               &    &VLA    & 3700 & 0.25 & $9.9\times10^{-4}$  & $86550\pm20$       & $93\pm2$       & -      & \cite{2021ApJ...908L..10H}\\
               &    &VLA    & 3200 & 0.25 & $2.4\times10^{-3}$  & $86550\pm20$       & $86\pm1$       & -      & \cite{2021ApJ...908L..10H}\\
 FRB~20171019A & yes  &GBT   & 820  & 0.10 & -                   & -         & -              & -      & \cite{kumar19}\\
 FRB~20180301A & yes &FAST   & 1378 & 0.12 & $3.4\times10^{-6}$  & $535\pm3$\tnote{*} & $85\pm1$    & -      & \cite{luo2020}\\
    &  &FAST  & 1088 & 0.12 & $1.5\times10^{-5}$  & $548\pm4$ & $61\pm4$    & -      & \cite{luo2020}\\
    &  &FAST  & 1047 & 0.12 & $1.7\times10^{-5}$  & $523\pm9$ & $40\pm3$    & -      & \cite{luo2020}\\
    &  &FAST  & 1160 & 0.12 & $1.0\times10^{-5}$  & $557\pm9$ & $60\pm9$   & -      & \cite{luo2020}\\
    &  &FAST   & 1094 & 0.12 & $1.5\times10^{-5}$  & $559\pm3$ & $70\pm2$   & -      & \cite{luo2020}\\
    &  &FAST   & 1051 & 0.12 & $1.8\times10^{-5}$  & $549\pm7$ & $59\pm5$    & -      & \cite{luo2020}\\
    &  &FAST   & 1441 & 0.12 & $2.7\times10^{-6}$  & $553\pm13$ & $85\pm5$   & -      & \cite{luo2020}\\
 FRB~20180916B & yes &CHIME  & 550  & 0.39 & $2.3\times10^{-4}$  & $-114.6 \pm0.6$    & $95\pm4$       & -      & \cite{chime8}  \\
    &   &LOFAR               & 165  & 0.01 & $5.0\times10^{-4}$  & $-115.71 \pm0.03$  & $70\pm4$       & -      & \cite{lofar21} \\
    &  &LOFAR               & 155  & 0.01 & $7.4\times10^{-4}$  & $-114.78 \pm0.09$  & $60\pm9$       & -      & \cite{lofar21} \\
    &  &LOFAR               & 115  & 0.01 & $4.4\times10^{-3}$  & $-114.43 \pm0.04$  & $30\pm4$       & -      & \cite{lofar21} \\
  FRB~20190303A & yes &CHIME  & 600  & 0.39 & $4.5\times10^{-3}$  & $-504.4\pm0.4$     & 21\tnote{$\dagger$}    & -      & \cite{chime9}\\
  FRB~20190604A & yes &CHIME   & 560  & 0.39 & $4.5\times10^{-6}$  & $-16\pm1$          & $100\pm10$     & -  & \cite{chime9} \\
  FRB~20190711A & yes &ASKAP  & 1220 & 4    & $1.4\times10^{-6}$  & $9\pm2$            & $101\pm2$\tnote{$\ddagger$} & $-1\pm2$\tnote{$\S$} & \cite{askap20}\\ 
\hline
  FRB~20110523A & no  &GBT & 800  & 0.05 & $1.8\times10^{-6}$  & $-186.1\pm1.4$     & $44\pm3$              & $23\pm30$      & \cite{masui2015}\\
  FRB~20140514A & no  &Parkes & 1400 & 0.39 & -                   & -                  & $0\pm10$\tnote{$\rVert$}     & $21\pm7$       & \cite{EP15}\\
 FRB~20150215A & no   &Parkes & 1400 & 0.39 & $4.4\times10^{-10}$ & $2\pm11$           & $43\pm5$              & $3\pm1$        & \cite{EP17}\\
 FRB~20150418A & no   &Parkes & 1400 & 0.39 & $1.4\times10^{-7}$  & $36\pm52$\tnote{$\P$} & $8.5\pm1.5$           & $0\pm4.5$      & \cite{Keane16}\\
 FRB~20150807A & no   &Parkes & 1400 & 0.39 & $1.6\times10^{-8}$  & $12\pm1$           & $80\pm1$              & $6\pm1$        & \cite{ravi16}\\
 FRB~20151230A & no   &Parkes & 1400 & 0.39 & -                   & -                  & $35\pm13$\tnote{$\P$}    & $6\pm11$       & \cite{caleb18}\\
 FRB~20160102A & no   &Parkes & 1400 & 0.39 & $5.3\times10^{-6}$  & $-221\pm6$         & $84\pm15$             & $30\pm11$      & \cite{caleb18}\\
 FRB~20180924B & no   &ASKAP & 1300 & 4    & $8.7\times10^{-6}$  & $22\pm2$           & $90.2\pm2.0$          & $-13.3\pm1.4$  & \cite{askap20}\\ 
 FRB~20181112A & no   &ASKAP & 1300 & 4    & $2.0\times10^{-6}$  & $10.5\pm4$         & 92\tnote{\#}           & -34\tnote{\#,**} & \cite{cho20}\\
 FRB~20190102C & no   &ASKAP & 1300 & 4    & $2.0\times10^{-4}$  & $-105\pm1$         & $82.2\pm0.7$\tnote{$\dagger\dagger$} & $4.8\pm0.5$    & \cite{askap20}\\
 FRB~20190608B & no   &ASKAP & 1300 & 4    & $2.2\times10^{-3}$  & $353\pm2$          & $91\pm3$              & $-9\pm2$       & \cite{askap20}\\
 FRB~20190611B & no   &ASKAP & 1300 & 4    & $7.2\times10^{-6}$  & $20\pm4$           & $70\pm3$\tnote{$\dagger\dagger$}     & $57\pm3$       & \cite{askap20}\\ 
\bottomrule

\end{longtable}
\end{scriptsize}
\end{ThreePartTable}
\end{center}

\clearpage

\begin{center}
\begin{ThreePartTable}
\begin{TableNotes}
  \item [*] {Ref~\cite{li21}.}
  \item [$\dagger$] {Ref~\cite{2021ApJS..257...59A}}
  \item [$\ddagger$] {Ref~\cite{luo2020}.}
  \item [$\S$] {Ref~\cite{chime8}.}
  \item [$\rVert$] {Ref~\cite{lofar21}.}  
  \item [\P] {Ref~\cite{chime9}.}
  \item [\#] {Ref~\cite{niu21}.}
  \item [**] {Ref~\cite{2021ATel14497....1C}.}
\end{TableNotes}
\begin{small}
\begin{longtable}{ccccccc}
  \rmfamily\\
  \caption{{\bf Properties of repeaters with $\sigma_{\mathrm{RM}}$ measurements.  } Column (1): name of the FRB;  Col.(2): Right Ascension (J2000); Col.(3): Declination (J2000), RAs and Decs are taken from the Transient Name Server\cite{tns}; Col.(4): dispersion measure; Col.(5): rotation measure, for FRBs~20121102A, 20180301A and 20180916B, we used averaged RM in Table~\ref{tab:data}, and for the others, we used averaged RM in Table~\ref{tab:burst}; Col.(6): RM scatter; Col.(7): scattering timescale, scaled to 1300\,MHz assuming $\tau_{\mathrm{sca}}\propto \lambda^4$, where $\lambda$ is the wavelength. `-' represents no reliable measurement.}\label{tab:frb_property}\\
  \toprule
 Name & RA & Dec & DM              &RM      & $\sigma_{\mathrm{RM}}$  & $\tau_{\mathrm{sca}}$ \\
      &             &     &(pc\,cm$^{-3}$)  & (rad m$^{-2}$)  & (rad m$^{-2}$)          & ms \\
  \midrule
  \endfirsthead

  \toprule
  \endhead
  
  \bottomrule
  \endfoot

  \bottomrule
  \insertTableNotes
  \endlastfoot

  FRB~20121102A   & 05$^h$32$^m$  & +33$^\circ$05'  & 565.8\tnote{*}& 81542   &  $30.9\pm0.4$   & $<$0.43\tnote{$\dagger$}           \\
  FRB~20180301A   & 06$^h$13$^m$  & +04$^\circ$39'  & 517\tnote{$\ddagger$} &546        &  $6.3\pm0.4$    & -              \\
  FRB~20180916B   & 01$^h$58$^m$  & +65$^\circ$44'  & 349.2\tnote{$\S$} &-115     &  $0.12\pm0.01$   & 0.009  \tnote{$\rVert$}        \\
  FRB~20190303A   & 13$^h$52$^m$  & +48$^\circ$07'  & 222.4\tnote{$\P$}&-411     &  $3.6\pm0.1$    & $0.19\pm0.04$\tnote{$\dagger$} \\
  FRB~20190417A   & 19$^h$39$^m$  & +59$^\circ$19'  & 1378.2\tnote{$\P$}&4681    &  $6.1\pm0.5$    & $0.21\pm0.06$ \tnote{$\dagger$}\\
  FRB~20190520B   & 16$^h$02$^m$  & -11$^\circ$17'  & 1210.3\tnote{\#}& 2759     &  $218.9\pm10.2$ & $9.8\pm2.0$ \tnote{\#} \\
  FRB~20201124A   & 05$^h$08$^m$  & +26$^\circ$03'  & 413.5\tnote{**}&-684     &  $2.5\pm0.1$    & 0.59  \tnote{$\dagger$}        \\
  
  \bottomrule

\end{longtable}
\end{small}
\end{ThreePartTable}
\end{center}

\begin{center}
\begin{ThreePartTable}
\begin{footnotesize}
\begin{longtable}{cccccccc}
  \rmfamily\\
  \caption{{\bf Polarization Properties of FRBs.} Column (1): burst index;  Col.(2): Modified Julian dates referenced to infinite frequency at the Solar System barycentre\cite{2006MNRAS.369..655H}; Col.(3): frequency of the burst weighted by signal to noise ratio.; Col.(4): fractional reduction in the linear polarization amplitude; Col.(5): RM obtained by RM-synthesis; Col.(6): RM obtained by Stokes QU-fitting; Col.(7): degree of linear polarization; Col.(8): degree of circular polarization. `-' represents not applicable.}\label{tab:burst}\\
  \toprule
Burst & MJD  & Frequency & $f_{\rm depol}$ & $\mathrm{RM_{FDF}}$ & $\mathrm{RM_{QUfit}}$ & \%~Linear & \%~Circular \\
      &                                & (MHz)       &                 & (rad m$^{-2}$)               & (rad m$^{-2}$)                 &           &             \\
  \midrule
  \endfirsthead

  \toprule
  \endhead
  
  \bottomrule
  \endfoot

  \bottomrule
  \endlastfoot
 
\multicolumn{4}{l}{FRB~20190520B (GBT 4.0-8.0\,GHz, $\Delta\nu$=0.37\,MHz)} \\
\hline
1 & 59292.45378759   & 4820 & $3.5\times10^{-7}$ & 2448$\pm$ 194   & $2168_{-49}^{+75}$   & $43\pm10$ & $-33\pm10$ \\
2 & 59296.43479679   & 4920 & $5.6\times10^{-7}$ & 3270$\pm$ 87    & $3250_{-71}^{+77}$   & $24\pm4$  & $-1\pm4$   \\
3 & 59300.46993279   & 5510 & $1.7\times10^{-7}$ & 2559$\pm$ 147   & $2877_{-159}^{+126}$ & $43\pm3$  & $9\pm3$    \\
\toprule
\multicolumn{4}{l}{FRB~20190303A (FAST 1.0-1.5\,GHz, $\Delta\nu$=0.12\,MHz)}
\\
\hline
1 & 59232.95530739  & 1370 & $1.9\times10^{-6}$ & $-$395 $\pm$ 14 & $-395_{-3}^{+2}$    & $96\pm16$ & $10\pm12$  \\
2 & 59258.95048596  & 1200 & $4.7\times10^{-6}$ & $-$416 $\pm$ 10 & $-444_{-5}^{+5}$     & $73\pm14$ & $3\pm11$   \\
3 & 59258.96114751  & 1320 & $2.7\times10^{-6}$ & $-$421 $\pm$ 11 & $-438_{-4}^{+7}$     & $96\pm11$ & $0\pm8$    \\
\toprule
\multicolumn{4}{l}{FRB~20190417A (FAST 1.0-1.5\,GHz, $\Delta\nu$=0.12\,MHz)} \\
\hline
1 & 59078.63698991   & 1128 & $8.8\times10^{-4}$ & 4755$\pm$ 7   & $4747_{-3}^{+4}$   & $64\pm5$ & $18\pm4$ \\
2 & 59078.65456755   & 1346 & $2.9\times10^{-4}$ & 4652$\pm$ 10   & $4655_{-2}^{+5}$   & $86\pm4$ & $9\pm3$ \\
3 & 59078.66719087   & 1125 & $8.4\times10^{-4}$ & 4614$\pm$ 7   & $4660_{-7}^{+6}$   & $69\pm10$ & $6\pm9$ \\
4 & 59078.66766456   & 1099 & $9.9\times10^{-4}$ & 4671$\pm$ 7    & $4670_{-5}^{+7}$   & $76\pm5$  & $10\pm4$   \\
5 & 59078.67199480   & 1075 & $1.2\times10^{-3}$ & 4711$\pm$ 16   & $4710_{-5}^{+5}$ & $52\pm7$  & $14\pm6$    \\
\toprule
\multicolumn{4}{l}{FRB~20201124A (FAST 1.0-1.5\,GHz, $\Delta\nu$=0.12\,MHz)} \\
\hline
1 & 59316.31897656   & 1121 & $2.0\times10^{-5}$ & -703 $\pm$ 1 & -703 $\pm$ 3     & $97\pm1$ & $-6\pm1$   \\
2 & 59316.33825543   & 1098 & $2.4\times10^{-5}$ & -730 $\pm$ 1 & -735 $\pm$ 4    & $99\pm1$ & $1\pm1$    \\
3 & 59316.34537882   & 1203 & $1.3\times10^{-5}$ & -710 $\pm$ 6 & -709 $\pm$ 1    & $98\pm1$ & $1\pm1$    \\
4 & 59316.34677951   & 1182 & $1.5\times10^{-5}$ & -717 $\pm$ 1 & -714 $\pm$ 4    & $95\pm1$ & $5\pm1$    \\
5 & 59316.34798014   & 1302 & $8.0\times10^{-6}$ & -697 $\pm$ 1 & -692 $\pm$ 2    & $98\pm1$ & $-4\pm1$    \\
6 & 59316.35036991   & 1122 & $2.0\times10^{-5}$ & -701 $\pm$ 1 & -703 $\pm$ 4    & $96\pm1$ & $-1\pm1$    \\
7 & 59316.38275891   & 1059 & $2.7\times10^{-5}$ & -685 $\pm$ 3 & -677 $\pm$ 4    & $97\pm2$ & $-7\pm1$    \\
8 & 59316.39130240   & 1070 & $2.6\times10^{-5}$ & -696 $\pm$ 1 & -700 $\pm$ 4    & $95\pm1$ & $-6\pm1$    \\
9 & 59316.39532239   & 1107 & $2.1\times10^{-5}$ & -697 $\pm$ 1 & -698 $\pm$ 4    & $99\pm1$ & $-6\pm1$    \\
10 & 59316.39727860  & 1165 & $1.6\times10^{-5}$ & -710 $\pm$ 1 & -707 $\pm$ 3    & $97\pm1$ & $-3\pm1$    \\
11 & 59316.39841730  & 1250 & $1.0\times10^{-5}$ & -688 $\pm$ 1 & -688 $\pm$ 4    & $98\pm1$ & $-1\pm1$    \\
\toprule
\multicolumn{4}{l}{FRB~20201124A (GBT 720-920\,MHz, $\Delta\nu$=0.20\,MHz)} \\
\hline
1 & 59315.03055957  & 840 & $2.7\times10^{-4}$ & -665$\pm$ 4& -662$\pm$ 1      & $80\pm3$  & $5\pm3$   \\
2 & 59315.04968094  & 842 & $2.2\times10^{-4}$ & -603$\pm$ 2& -603$\pm$ 1    & $76\pm3$  & $26\pm2$    \\
3 & 59315.05857900  & 844 & $3.0\times10^{-4}$ & -707$\pm$ 22& -702$\pm$ 2    & $75\pm3$  & $-21\pm2$    \\
4 & 59315.07106791  & 836 & $2.9\times10^{-4}$ & -677$\pm$ 3& -681$\pm$ 1    & $75\pm3$  & $-6\pm2$    \\
5 & 59315.07666824  & 839 & $2.4\times10^{-4}$ & -623$\pm$ 2& -621$\pm$ 2    & $88\pm1$  & $-5\pm1$    \\
6 & 59315.07878012  & 820 & $2.9\times10^{-4}$ & -634$\pm$ 1& -640$\pm$ 1   & $81\pm3$  & $-10\pm2$    \\
7 & 59315.08175219  & 850 & $2.8\times10^{-4}$ & -698$\pm$ 3& -701$\pm$ 1    & $83\pm2$  & $-15\pm2$    \\
8 & 59315.08311698  & 832 & $3.0\times10^{-4}$ & -679$\pm$ 2& -682$\pm$ 1    & $78\pm2$  & $-2\pm2$    \\
9 & 59315.08562879  & 844 & $2.7\times10^{-4}$ & -674$\pm$ 1& -674$\pm$ 1    & $82\pm3$  & $-15\pm2$    \\
\toprule
\multicolumn{4}{l}{FRB~20121102A (FAST 1.0-1.5\,GHz, $\Delta\nu$=0.12\,MHz)} \\
\hline
- & -  & 1250 & $2.0\times10^{-1}$ & -& -      & $<$6  & -   \\
\toprule
\multicolumn{4}{l}{FRB~20190520B (FAST 1.0-1.5\,GHz, $\Delta\nu$=0.12\,MHz)} \\
\hline
- & -  & 1250 & $1.9\times10^{-4}$ & -& -      & $<$20  & -   \\
\bottomrule
\end{longtable}
\end{footnotesize}
\end{ThreePartTable}
\end{center}

\clearpage
\begin{flushleft}
\justifying
\end{flushleft}